\documentclass[prb,11pt,onecolumn,english,superscriptaddress,nofootinbib]{revtex4-2}

\usepackage{epsfig,bm,epsf,graphicx,float}
\usepackage{amsmath,amssymb,mathrsfs,amsthm,mathtools}
\usepackage{bm,bbm}
\usepackage{graphics,graphicx}
\graphicspath{{images/}}
\usepackage{lmodern}
\usepackage{color,xcolor}
\usepackage{rotating}
\usepackage{subfigure}
\usepackage{hyperref}
\usepackage{enumerate}
\usepackage{comment}
\usepackage[textsize=footnotesize,textwidth=1.5in,disable]{todonotes}

\usepackage{xfrac}

\newcommand{\bra}[1]{\langle#1\!|}
\newcommand{\ket}[1]{|\!#1\rangle}

\newcommand{\wket}[1]{|#1\rangle}
\newcommand{\braket}[2]{\langle #1|#2\rangle}
\newcommand{\down}{\downarrow}
\newcommand{\id}{\mathbbm{1}}
\newcommand{\bpm}{\begin{pmatrix}}
\newcommand{\epm}{\end{pmatrix}}
\newcommand{\bmm}{\begin{matrix}}
\newcommand{\emm}{\end{matrix}}
\newcommand{\up}{\uparrow}

\newcommand{\bea}{\begin{eqnarray}}
\newcommand{\eea}{\end{eqnarray}}

\newcommand{\op}[1]{\operatorname{#1}}

\usepackage{tikz}
\usepackage{pgfplotstable}
\usetikzlibrary{shapes,positioning,arrows,calc,intersections,through}
\tikzstyle{vtx}=[circle,fill=black,inner sep=0pt,minimum size=.7mm]

\begin{document}
\title{Order, Disorder, and Transitions in Decorated AKLT States on Bethe
Lattices}

\author{Nicholas Pomata}
\email{nicholas.pomata@stonybrook.edu}
\affiliation{C. N. Yang Institute for Theoretical Physics and Department of
Physics and Astronomy, State University of New York at Stony Brook, Stony
Brook, NY 11794-3840, USA}
\date{today}

\begin{abstract}
Returning to one of the original generalizations of the AKLT state, we extend  
prior analysis on the Bethe lattice (or Cayley tree) to a variant with
a series of $n$ spin-1 decorations placed on each edge. The recurrence relations
derived for this system demonstrate that such systems are ``critical'' for
coordination numbers $z=3^{n+1}+1$, demonstrating order for greater and
disorder for lesser coordination number. We then generalize further, effectively
interpolating between systems with different values of $n$, using two
realizations, one isotropic under local $SU(2)$ transformations and one
anisotropic. Exact analysis of these recurrence relations allows us to deduce
the location and behavior of order-disorder phase transitions for $z>4$.
\end{abstract}

\maketitle

\section{Introduction}
The valence-bond states developed by Affleck, Kennedy, Lieb, and Tasaki
\cite{AKLT_PRL} (AKLT), foundational for the tensor-network formalism, provided
a useful framework to explore strongly-interacting systems. The one-dimensional
states originally explored are particularly amenable to study due to the
transfer-matrix analysis they admit. However, much as with classical spin
models that admit a similar analysis in one dimension, AKLT's
higher-dimensional generalizations \cite{AKLT_1988} are largely quite difficult
to ascertain exact information about, effectively because loops in
higher-dimensional lattices obstruct such a transfer-matrix approach.
There is one notable exception: the Bethe lattice, or Cayley tree, an
infinitely-branching graph with an identical number of edges (the degree or
coordination number $z$) per site, on which AKLT formulated exact recurrence
relations that enabled them to determine exact properties of this system.
Following their work, Fannes, Nachtergaele, and Werner established more
thorough conclusions, analyzing the state using the quantum Markov chain
formalism to rigorously investigate the possible states on the infinite lattice
\cite{fannes1992cayley,werner1989construction}.

\begin{figure}
    %\tikzexternalenable
    %\tikzsetfigurename{bethepentagon}
    \begin{tikzpicture}[scale=2]
      \clip (0,0) circle (1);
      \pgfplotstableread[col sep=comma]{tikzfigs/crcpent4.csv}\penttable
      \pgfplotstableforeachcolumnelement{radius}\of\penttable\as\rad{
        \pgfplotstablegetelem{\pgfplotstablerow}{xcoord}\of{\penttable}
        \pgfmathsetmacro\xco{\pgfplotsretval}
        \pgfplotstablegetelem{\pgfplotstablerow}{ycoord}\of{\penttable}
        \pgfmathsetmacro\yco{\pgfplotsretval}
        \draw[blue,ultra thin] (\xco,\yco) circle (\rad);
      }
      \draw[blue,very thin] (-1,0) -- (1,0);
      \draw[blue,very thin] (0,-1) -- (0,1);
      \pgfplotstableread[col sep=comma]{tikzfigs/crcbethe4.csv}\bethetable
      \pgfplotstableforeachcolumnelement{radius}\of\bethetable\as\rad{
        \pgfplotstablegetelem{\pgfplotstablerow}{xcoord}\of{\bethetable}
        \pgfmathsetmacro\xco{\pgfplotsretval}
        \pgfplotstablegetelem{\pgfplotstablerow}{ycoord}\of{\bethetable}
        \pgfmathsetmacro\yco{\pgfplotsretval}
        \draw[blue,very thin] (\xco,\yco) circle (\rad);
      } 
      \draw[red,semithick] (-1,0) -- (1,0);
      \draw[red,semithick] (0,-1) -- (0,1);
      \pgfplotstableread[col sep=comma]{tikzfigs/crcbethe4.csv}\bethetable
      \pgfplotstableforeachcolumnelement{radius}\of\bethetable\as\rad{
        \pgfplotstablegetelem{\pgfplotstablerow}{xcoord}\of{\bethetable}
        \pgfmathsetmacro\xco{\pgfplotsretval}
        \pgfplotstablegetelem{\pgfplotstablerow}{ycoord}\of{\bethetable}
        \pgfmathsetmacro\yco{\pgfplotsretval}
        \draw[red,semithick] (\xco,\yco) circle (\rad);
      } 
    \end{tikzpicture}
    %\tikzsetfigurename{betheapeir}
    \begin{tikzpicture}[scale=2,even odd rule]
      \clip (0,0) circle (1);
      \fill[red] (0,0) circle (1);
      \pgfplotstableread[col sep=comma,header=has colnames]{tikzfigs/crcapeir4.csv}\bethetable
      \pgfpathmoveto{\pgfpoint{-1cm}{0cm}}
      \pgfpathlineto{\pgfpoint{1cm}{0cm}}
      \pgfpatharc{0}{180}{1cm}
      \pgfpathmoveto{\pgfpoint{0cm}{-1cm}}
      \pgfpathlineto{\pgfpoint{0cm}{1cm}}
      \pgfpatharc{90}{270}{1cm}
      \pgfplotstableforeachcolumnelement{radius}\of\bethetable\as\rad{
        \pgfplotstablegetelem{\pgfplotstablerow}{xcoord}\of{\bethetable}
        \pgfmathsetmacro\xco{\pgfplotsretval}
        \pgfplotstablegetelem{\pgfplotstablerow}{ycoord}\of{\bethetable}
        \pgfmathsetmacro\yco{\pgfplotsretval}
        %(\xco,\yco) circle (\rad)
        \pgfpathcircle{\pgfpoint{\xco cm}{\yco cm}}{\rad cm}
      }
      \pgfsetfillcolor{blue!30}
      \pgfusepath{fill}
    \end{tikzpicture}
  %\tikzexternaldisable
  \caption{The Bethe lattice can be embedded in hyperbolic geometry in regular
  ways: e.g. as (a) a sublattice of a more conventional tiling (here the
  degree-4 Bethe lattice within the order-4 pentagonal tiling) or as (b) the
  edge set of a tiling by infinite-sided shapes, or apierogons.}
  \label{fig:Bethe-hyperbolic}
\end{figure}
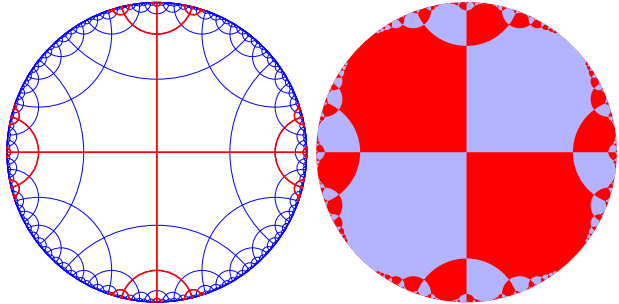

The Bethe lattice, while generally unphysical due to its exponentially-growing
vertices\footnote{Because of this, the Bethe lattice cannot be embedded in
Euclidean space with both strictly finite node density and bounded edge
length. However, as shown in Fig.~\ref{fig:Bethe-hyperbolic}, it can be
embedded in hyperbolic space in interesting, if not necessarily physically
relevant, ways.}, has attracted attention for use as a test bed for tensor
network systems, especially infinite tensor-network systems with the develoment
of the quantum Markov chain formalism to rigorously
define infinite tensor-network states \cite{nachtergaele1990qmarkov,
fidaleo-mukhamedov2005factors,accardi-mukhamedov2016construction},
and as a kind of mean-field theory variant. This includes analysis of
Hamiltonians on the Bethe lattice without exact ground states, using
variants of the density-matrix renormalization group \cite{friedman1997dmrg}
and the time-evolving block decimation algorithm \cite{nagaj2008mps} to
extract approximate ground states and analyze their phase diagrams.

In this work, we will return to the AKLT state on the Bethe lattice, this
time adding decorations on each edge as was done in
\cite{decorated1,decorated2} to make two-dimensional AKLT systems more
amenable to study. In Section~\ref{sec:recurrence}, we will extend the methods
of \cite{AKLT_1988} to these systems and obtain exact recurrence
relations, including indications of critical behavior in some cases.
In Section~\ref{sec:loop}, we will describe observables that can be used
to validate the idea of such long-ranged behavior on an infinite disordered
state. In Section~\ref{sec:interpolating}, we will then describe continuous
extensions of these decorated systems and show that these exhibit phase
transitions with critical features that can be described exactly.
Finally, in Section~\ref{sec:anisotropic}, we will examine the somewhat
complicated phase diagram that occurs from this when the system explicitly
breaks $SU(2)$ symmetry.

\section{Basic recurrence relations}
\label{sec:recurrence}
In determining the behavior of the AKLT model on decorated lattices, we will,
for the most part, follow the derivation given by AKLT \cite{AKLT_1988} of the
behavior of the model on the ``bare'' Bethe lattice. In particular, we will
start by considering finite systems with antiferromagnetic boundary conditions;
we will then extend that analysis to make tentative conclusions about the
the systems as may be defined in infinite-size frameworks where the translation
symmetry of the Bethe lattice is possible.

The derivation starts with the expression for the norm-squared of an AKLT state 
$\Psi_A$ on a bipartite graph $A$ (from the equation shortly before (4.8) of
\cite{AKLT_1988})
\begin{equation}
\braket{\Psi_A}{\Psi_A} = \sum_{G,G'}\prod_{i\in A}\delta(m_i(G),m_i(G'))m_i(G)!(z_i-m_i(G))!.
\label{eq:norm-loops}
\end{equation}
Here
\begin{itemize}
\item $G$ and $G'$ are subgraphs (alternatively, edge subsets) of $A$,
\item $z_i$ is the connectivity of vertex $i$ in the graph $A$ (which equals
$z$ on the sites of the bare lattice and 2 on the decorations),
\item $m_i(G)$ is the connectivity of vertex $i$ in the graph $G$, and
\item $\delta$ is the Kronecker delta symbol.
\end{itemize}
For the sake of brevity, we will for the remainder of this section use the
function
\[ c_z(n) \equiv n!(z-n)!=\frac{z!}{\binom{z}{n}}. \]

We may interpret $n_i$ as the physical index of spin $i$ in the $S_z$ basis
\textit{on every other site of} $A$; that is, if we bipartition $A$ with
some $\op{sgn}(i)$ that maps alternating sites to $+1$ and $-1$, we can
identify the $S_z$ index of \textit{any} spin $i$ with
$z/2+\op{sgn}(i)(m_i-z/2)$.
In particular, we can insert a matrix $M_{ab}$ contracted with the physical
indices at site $i$ by modifying
\begin{align*}
\delta(n_i(G),&m_i(G'))c_{z_i}(m_i(G))\Rightarrow\\
&\sqrt{c_{z_i}(m_i(G))c_{z_i}(m_i(G'))}\\&\times
\left\{\begin{array}{cr}
M_{m_i(G),m_i(G')}&\op{sgn}(i)=+1\\
M_{z_i-m_i(G),z_i-m_i(G')}&\op{sgn}(i)=-1
\end{array}\right..
\end{align*}

When considering a finite system, we will within this section use what we
may call ``classical'' boundary conditions, in which boundary qubits are fixed
to be either spin-up or spin-down. That is, when considering a system with
radius $M$, we eliminate the physical spins at a distance $M+1$ or more
from the origin and replace the virtual singlets connecting sites of distance
$M$ and $M+1$ from the origin with up or down spins. (This is the only sort
of configuration considered in \cite[, shortly before 4.8]{AKLT_1988}). In that case, the 
only admissible configurations in \eqref{eq:norm-loops} are ones in which
$G$ and $G'$ coincide exactly, i.e. as in \cite[, eq. 4.8]{AKLT_1988},
the equation reduces to
\begin{equation}
\braket{\Psi_A}{\Psi_A} = \sum_{G\subset A}\prod_{i\in A}c_{z_i}(m_i(G)).
\label{eq:norm-gen}
\end{equation}

\begin{figure}
  %\tikzsetfigurename{bethebranches}
  \subfigure[]{
  \begin{tikzpicture}[scale=2.5]
    \clip (0,0) circle (1);
    \pgfplotstableread[col sep=comma]{tikzfigs/crcapeir4.csv}\circtable
    \pgfplotstableforeachcolumnelement{radius}\of\circtable\as\rad{
      \pgfplotstablegetelem{\pgfplotstablerow}{xcoord}\of{\circtable}
      \pgfmathsetmacro\xco{\pgfplotsretval}
      \pgfplotstablegetelem{\pgfplotstablerow}{ycoord}\of{\circtable}
      \pgfmathsetmacro\yco{\pgfplotsretval}
      \draw (\xco,\yco) circle (\rad);
    }
    \node[vtx,minimum size=2] (origin) at (0,0) {};
    \pgfplotstableread[col sep=comma]{tikzfigs/vtxapeir4.csv}\ptstable
    \pgfplotstableforeachcolumnelement{id}\of\ptstable\as\label{
      \pgfplotstablegetelem{\pgfplotstablerow}{xcoord}\of{\ptstable}
      \pgfmathsetmacro\xco{\pgfplotsretval}
      \pgfplotstablegetelem{\pgfplotstablerow}{ycoord}\of{\ptstable}
      \pgfmathsetmacro\yco{\pgfplotsretval}
      \node[vtx,minimum size=.6mm] (\label) at (\xco,\yco) {};
    }
    \draw[thick,dashed] (0) -- (2);
    \draw[thick,dashed] (1) -- (3);
    \draw (0) -- (1,0);
    \draw (1) -- (0,1);
    \draw (2) -- (-1,0);
    \draw (3) -- (0,-1);
    \node (Blabel) at (.63,.07) {$B_6$}; 
    \clip (0,0) -- (1,1) -- (1,-1) -- (0,0);
    \pgfplotstableforeachcolumnelement{id}\of\ptstable\as\label{
      \draw[blue!80,radius=.15mm,inner sep=0] (\label) circle; 
    }
    \node (Bsublabel) at (.74,.35) {$B_5$};
  \end{tikzpicture}
  }
  \subfigure[]{
  \begin{tikzpicture}[scale=2.5]
    \clip (0,0) circle (1);
    \node (Y6label) at (.15,.0) {$Y_6=\sum$}; 
    \node[red] (B5a) at (.74,.32) {$Y_5$};
    \node[green] (B5b) at (.82,.08) {$Z_5$};
    \node[green] at (.74,-.32) {$Z_5$};
    \clip (.4,-.7) rectangle (1,.7);
    \pgfplotstableread[col sep=comma]{tikzfigs/crcapeir4.csv}\circtable
    \pgfplotstableforeachcolumnelement{radius}\of\circtable\as\rad{
      \pgfplotstablegetelem{\pgfplotstablerow}{xcoord}\of{\circtable}
      \pgfmathsetmacro\xco{\pgfplotsretval}
      \pgfplotstablegetelem{\pgfplotstablerow}{ycoord}\of{\circtable}
      \pgfmathsetmacro\yco{\pgfplotsretval}
      \draw (\xco,\yco) circle (\rad);
    }
    \node[vtx,minimum size=2] (origin) at (0,0) {};
    \pgfplotstableread[col sep=comma]{tikzfigs/vtxapeir4.csv}\ptstable
    \pgfplotstableforeachcolumnelement{id}\of\ptstable\as\label{
      \pgfplotstablegetelem{\pgfplotstablerow}{xcoord}\of{\ptstable}
      \pgfmathsetmacro\xco{\pgfplotsretval}
      \pgfplotstablegetelem{\pgfplotstablerow}{ycoord}\of{\ptstable}
      \pgfmathsetmacro\yco{\pgfplotsretval}
      \node[vtx,minimum size=.6mm] (\label) at (\xco,\yco) {};
    }
    \draw[red] (0) -- (2);
    \draw (0) -- (1,0);
    \draw (1) -- (0,1);
    \draw (2) -- (-1,0);
    \draw (3) -- (0,-1);
    \clip (0,0) -- (1,1) -- (1,-1) -- (0,0);
    \pgfplotstableforeachcolumnelement{id}\of\ptstable\as\label{
      \draw[blue!80,radius=.15mm,inner sep=0] (\label) circle; 
    }
  \end{tikzpicture}
  }
  \subfigure[]{
  \begin{tikzpicture}[scale=2.5]
    \clip (0,0) circle (1);
    \node (Z5label) at (.15,.0) {$Z_5=\sum$}; 
    \node[green] (B5a) at (.74,.32) {$Z_4$};
    \node[red] (B5b) at (.82,.08) {$Y_4$};
    \node[green] (B5c) at (.74,-.32) {$Z_4$};
    \clip (.4,-.7) rectangle (1,.7);
    \pgfplotstableread[col sep=comma]{tikzfigs/crcapeir4.csv}\circtable
    \pgfplotstableforeachcolumnelement{radius}\of\circtable\as\rad{
      \pgfplotstablegetelem{\pgfplotstablerow}{xcoord}\of{\circtable}
      \pgfmathsetmacro\xco{\pgfplotsretval}
      \pgfplotstablegetelem{\pgfplotstablerow}{ycoord}\of{\circtable}
      \pgfmathsetmacro\yco{\pgfplotsretval}
      \draw (\xco,\yco) circle (\rad);
    }
    \node[vtx,minimum size=2] (origin) at (0,0) {};
    \pgfplotstableread[col sep=comma]{tikzfigs/vtxapeir4.csv}\ptstable
    \pgfplotstableforeachcolumnelement{id}\of\ptstable\as\label{
      \pgfplotstablegetelem{\pgfplotstablerow}{xcoord}\of{\ptstable}
      \pgfmathsetmacro\xco{\pgfplotsretval}
      \pgfplotstablegetelem{\pgfplotstablerow}{ycoord}\of{\ptstable}
      \pgfmathsetmacro\yco{\pgfplotsretval}
      \node[vtx,minimum size=.6mm] (\label) at (\xco,\yco) {};
    }
    \draw[green] (0) -- (2);
    \draw (0) -- (1,0);
    \draw (1) -- (0,1);
    \draw (2) -- (-1,0);
    \draw (3) -- (0,-1);
    \clip (0,0) -- (1,1) -- (1,-1) -- (0,0);
    \pgfplotstableforeachcolumnelement{id}\of\ptstable\as\label{
      \draw[blue!80,radius=.15mm,inner sep=0] (\label) circle; 
    }
  \end{tikzpicture}
  }
  \caption{Factoring the expression \eqref{eq:norm-gen} for the norm of the
  AKLT state into ``branches'' of the Bethe lattice. (a) We label branches of a
  finite lattice (equivalent under ``rotations'') by the distance of the root
  node from the boundary. (b) The values $Y_M$ and $Z_M$, corresponding to
  inclusion in or exclusion from $G$, are determined from $Y_{M-1}$ and
  $Z_{M-1}$ given the possible configurations. (c) These in turn have been
  determined from $Y_{M-2}$ and $Z_{M-2}$.}
\label{fig:branch-factor}
\end{figure}
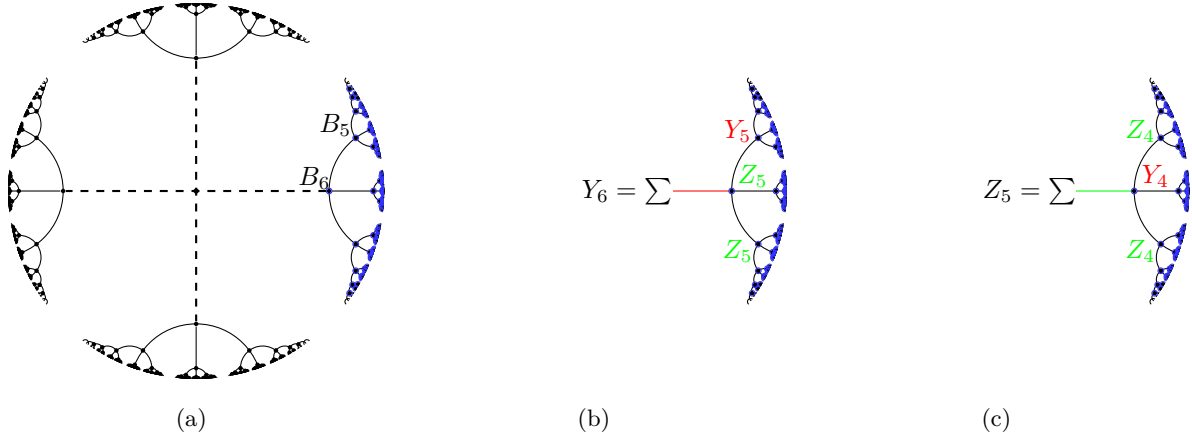

The recursive solution used relies on the ability to factor branches of the
``Cayley tree'' out of this expression, in what is a variant on a
transfer-matrix approach. Again as in \cite{AKLT_1988}, and as shown in
Fig.~\ref{fig:branch-factor}, we define $Y_M$ and $Z_M$ as the value of
the factor corresponding to a depth-$M$ branch $B_M$ when the edge that connects
that branch to the origin is, respectively, included in or excluded from $G$.
That is (if we consider $i=0$ to be the origin of the branch $B_M$),
\begin{align}
Y_M &\equiv \sum_{G\subset B_M}c_z(m_0(G)+1)\prod_{k\in B_M}c_{z_k}(m_k(G))
\label{eq:Yfull}\\
Z_M &\equiv \sum_{G\subset B_M}c_z(m_0(G))\prod_{k\in B_M}c_{z_k}(m_k(G))
\label{eq:Zfull}\\
\|\Psi_M\|^2 &= \sum_{m=0}^z\binom{z}{m}c_z(m)Y_M^m Z_M^{z-m}=z!\sum_{m=0}^z Y_M^mZ_M^{z-m}.\label{eq:norm-bare}
\end{align}
(We are, for now, only considering the bare lattice; while \eqref{eq:Yfull} and
\eqref{eq:Zfull} remain correct for the decorated lattice, \eqref{eq:norm-bare}
will need to be modified.)

In analyzing the bare lattice in \cite{AKLT_1988}, AKLT factor the sub-branches
of $B_M$ out as well to yield recurrence relations for $Y_M$ and $Z_M$:
\footnote{
Note that, since $Y_M$ and $Z_M$ are initially defined for a branch adjoining
the origin, i.e. with an ``odd'' root vertex at depth 1, using them in reference
to a branch with a root vertex at depth 2, by altering the parity of the
vertices, implicitly inverts the boundary conditions. We will rectify this by
letting the branch $B_M$ terminate in $\ket{\uparrow}$, and have an odd-parity
root vertex, if $M$ is even \textit{or the number of decorations $n$ is odd},
and terminate in $\ket{\downarrow}$ if $M$ is odd and $n$ is even.}
\todo{Check parity of $\up$/$\down$ and $M$}
\begin{equation}
\begin{split}
Y_M &= \sum_{m=0}^{z-1}(z-1)!(m+1)Y_{M-1}^mZ_{M-1}^{z-m-1}
\\
Z_M &= \sum_{m=0}^{z-1}(z-1)!(z-m)Y_{M-1}^{m}Z_{M-1}^{z-m-1}
\end{split}
\label{eq:recur-bare}
\end{equation}
We can modify this approach to suit the $n$-fold decorated lattice by treating
each decoration as the root of its own branch: that is, let $B^{(0)}_M=B_M$ be
a branch with a degree-$z$ vertex as its root (and a depth \textit{in
degree-$z$ vertices} of $M$) and $B^{(k)}_M$ be that branch with $k$
decorations. The primary modifications to \eqref{eq:norm-bare} and
\eqref{eq:recur-bare} consist in changing $z$ to 2 as appropriate:
\begin{equation}
\begin{split}
Y^{(0)}_M &= (z-1)!\sum\limits_{m=0}^{z-1}(m+1)(Y^{(n)}_{M-1})^m(Z_{M-1}^{(n)})^{z-m-1}\\
Y^{(k)}_M &= 2Y_{M}^{(k-1)} + Z_{M}^{(k-1)},\ k > 0\\
Z^{(0)}_M &= (z-1)!\sum\limits_{m=0}^{z-1}(z-m)(Y^{(n)}_{M-1})^m(Z_{M-1}^{(n)})^{z-m-1}\\
Z^{(k)}_M &= Y_{M}^{(k-1)} + 2Z_{M}^{(k-1)},\ k > 0
\end{split}
\label{eq:decoratedYZ}
\end{equation}
The $k>0$ case is a simple linear relation that we can solve as a matrix
equation:
\begin{align}
\left(\begin{array}{c} Y_M^{(n)}\\ Z_M^{(n)} \end{array}\!\right) 
&=
\left(\begin{array}{ccc} 2&&1 \\ 1&&2 \end{array}\right)
\left(\begin{array}{c} Y_M^{(n-1)}\\ Z_M^{(n-1)} \end{array}\!\right)
%\notag\\
%&=
=
\left(\begin{array}{ccc} 2&&1 \\ 1&&2 \end{array}\right)^{\textstyle n}
\left(\begin{array}{c} Y_M^{(0)}\\ Z_M^{(0)} \end{array}\!\right)\notag\\
&=
\left(\begin{array}{cc} \frac{3^n+1}{2}&\frac{3^n-1}{2} \\ \frac{3^n-1}{2}&\frac{3^n+1}{2} \end{array}\right)
\left(\begin{array}{c} Y_M^{(0)}\\ Z_M^{(0)} \end{array}\!\right).
\label{eq:transfer}
\end{align}

Combining \eqref{eq:decoratedYZ} with \eqref{eq:transfer}, we get
\begin{equation}
\begin{split}
Y_{M+1}^{(n)} &= \frac{(z-1)!}{2}\sum_{i=0}^{z-1}[(3^n+1)(i+1)+(3^n-1)(z-i)](Y^{(n)}_{M-1})^m(Z_{M-1}^{(n)})^{z-m-1} \\
Z_{M+1}^{(n)} &= \frac{(z-1)!}{2}\sum_{i=0}^{z-1}[(3^n-1)(i+1)+(3^n+1)(z-i)](Y^{(n)}_{M-1})^m(Z_{M-1}^{(n)})^{z-m-1} 
\end{split}
\end{equation}
We now set $W_M=Y_M^{(n)}/Z_M^{(n)}$, extending \cite[, eq. 4.9]{AKLT_1988} into
the recurrence relation
\begin{equation}
W_{M+1}=f_{n,z}(W_M)=\frac{\sum\limits_{i=0}^{z-1}\left(3^n(z+1)+2i+1-z\right)W_M^i}{\sum\limits_{i=0}^{z-1}\left(3^n(z+1)-2i-1+z\right)W_M^i}
\label{eq:recurnz}
\end{equation}
Accounting for all of the decorations on the inward-reaching edge allows us to
use, without modification, the equation in [, following 4.11]\cite{AKLT_1988}
for the polarization of the central spin: 
\begin{equation}
\frac{\langle\Psi_M|S_z^{(0)}|\Psi_M\rangle}{\braket{\Psi_M}{\Psi_M}} = \frac{\sum\limits_{i=0}^z(z/2-i)W_M^i}{\sum\limits_{i=0}^z W_M^i} \equiv m_z(W_M).
\label{eq:sz-exact}
\end{equation}
In particular, we note that $m_z$ is monotonic decreasing, with
$m_z(0)=z/2$, $m_z(1)=0$, and $m_z(\infty)=-z/2$.

\subsection{Consequences of the recurrence relation}
Much as observed by AKLT in the bare case, the basic key features of $f_{n,z}$
are that it is monotonic nondecreasing (for $0\leq W\leq \infty$) and that
$W=1$ is a fixed point. We also note that it is manifestly true from the
spin-flip-invariance of the system, which exchanges the variables $Y$ and $Z$
defined in \eqref{eq:Yfull}, \eqref{eq:Zfull}, that
$f_{n,z}(W^{-1})=f_{n,z}(W)^{-1}$. With the additional datum that
\[ f_{n,z}(0)=\frac{3^n(z+1)-z+1}{3^n(z+1)+z-1}>0,\]
we may conclude that:

\begin{enumerate}
\item Antiferromagnetic order corresponds to the existence of a fixed point of
$f_{n,z}$ at some $0\leq W_0<1$.
\item In order to establish that $W=1$ is the unique (nonnegative) fixed point 
of $f_{n,z}$, it is necessary and sufficient to prove that $f(W)>W$ for all
$0<W<1$.
\item When the fixed point $W=1$ is unstable, i.e. $f'(W)>1$,
it is necessarily true that $f(W)<W$ in some neighborhood $(1-\varepsilon,1)$,
and therefore there is at least one additional ordered, attractive fixed point.
\end{enumerate}
With our rudimentary analysis, therefore, we will generally be able to
conclusively prove the \textit{existence} of antiferromagnetic order
(without knowing the value of the fixed point, which corresponds to the
value of the magnetization) but, in cases believed to be disordered, we will
not be able to \textit{disprove} the existence of such a fixed point.

Therefore, our guide to the behavior of the system is
\begin{equation}
f'_{n,z}(1) = \frac{z-1}{3^{n+1}}.
\label{eq:fprime}
\end{equation}
This suggests the absence of order when $z < 3^{n+1}+1$ and implies the
presence of order when $z > 3^{n+1}+1$.

We now consider the marginal case $z = 3^{n+1}+1$.
As the denominator of \eqref{eq:recurnz} is positive-definite (with
a value of $3^nz(z+1)$ at $W=1$), we may determine
the leading-order behavior of $f_{n,z}(1-\varepsilon)-(1-\varepsilon)$
(for $\varepsilon\equiv 1-W$ small) expanding the numerator:
\begin{equation}
\begin{split}
\sum_{i=0}^{z-1}&\left(3^n(z+1) + 2i + 1 - z\right)(1-\varepsilon)^i - \sum_{i=0}^{z-1}\left(3^n(z+1) - 2i - 1 + z\right)(1-\varepsilon)^i\\
\simeq& \frac{z(z+1)}{3}(3^{n+1}-z+1)\varepsilon
- \frac{z(z^2-1)}{6}(3^{n+1}-z+1)\varepsilon^2
+ \frac{z(z^2-1)(z-2)}{60}(10\times 3^n - 3z + 4)\varepsilon^3
\end{split}
\label{eq:numorder3}
\end{equation}
In particular, when $z=3^{n+1}+1$ the first two terms vanish, and
(substituting for $(z-1)/3$ for $3^n$) the sub-leading-order behavior of
$f_{n,z}$ is
\begin{equation}
f_{n,z}(W)-W \simeq \frac{(1-W)^3}{60}(z^2-4) > 0
\label{eq:marginal-leading}
\end{equation}
for $1-W>0$ small.

\subsection{Asymptotic behavior}
We first state the leading-order behavior of \eqref{eq:sz-exact}:
\begin{equation}
 m_z(1-\varepsilon) \simeq \frac{1}{z+1}\sum_{i=0}^z\left(\frac{z}{2}-i\right)(1-i\varepsilon) 
%= \frac{1-W}{z+1}\left(\frac{z}{2}\sum_{i=0}^z i - \sum_{i=0}^z i^2\right)
%\notag\\
= \frac{z(z+2)}{12}\varepsilon.
\label{eq:S-order1}
\end{equation}
From \eqref{eq:fprime}, we can determine the asymptotic behavior of $W_M$ for
the disordered case $z<3^{n+1}+1$ (at least, assuming that there are
no unexpected fixed points):
\begin{align}
f_{n,z}(1-\varepsilon) &\simeq 1 - \frac{z-1}{3^{n+1}}\varepsilon\notag\\
1 - W_M &\sim \left(\frac{z-1}{3^{n+1}}\right)^M\notag\\
m_z(W_M) &\sim \left(\frac{z-1}{3^{n+1}}\right)^M.
\label{eq:mzdecay}
\end{align}
In other words, we may say there is an effective correlation length, of sorts,
$\xi^{-1} = (n+1)\ln 3 - \ln(z-1)$.

Moving to the marginal case $z=3^{n+1}+1$, we note the following: If
a function $g$ has leading-order behavior
$g(\varepsilon)= 1-C\varepsilon^r+\mathcal{O}(\varepsilon^{r+1})$ for some
coefficient $C$ and exponent $r \geq 2$, it can be shown that
\[ g\!\left(\frac{1}{\sqrt[r-1]{(r-1)CM}}\right) = \frac{1}{\sqrt[r-1]{(r-1)C(M+1)}} + \mathcal{O}(M^{-\frac{r}{r-1}}). \]
In particular, we may use the leading-order expansion in
\eqref{eq:marginal-leading}: for $g(1-W) \equiv 1 - f_{n,z}(W)$, which is
to say $r = 3$ and $C = \frac{z^2-4}{60}$ in the ansatz for $g$, we determine
that the large-radius behavior of our order parameter will be governed by
\begin{equation}
\begin{split}
W_M &\simeq 1 - \sqrt{\frac{30}{(z^2-4)M}}\\
m_z(W_M) &\simeq \frac{z}{2}\sqrt{\frac{z+2}{z-2}\times\frac{5}{6M}}.
\end{split}
\label{eq:mz-power-law}
\end{equation}
For the undecorated case, when $z=4$, this becomes
$m_z(W_M)\simeq \sqrt{\frac{10}{M}}$.

\subsection{Numerical confirmation}
Due to the inherent tensor-network ``valence bond state'' structure of
the AKLT state and the loop-free nature of the Bethe lattice, when $z$ is
sufficiently small it is easy to directly compute the value of the N\'eel order
parameter $m_z$ and other observables on finite-size subsystems. We do so as
follows:
\begin{enumerate}[i.]
  \item We take a ``boundary state'', a one-qubit density matrix
  $\rho_m$ initialized to $\rho_0=\ket{\uparrow}\bra{\uparrow}$.
  \item We apply a generalization of the transfer matrix for degree $z$,
  constructed by contracting the spin-$z/2$ projector along its physical
  index and then contracting the final\todo{check ``final'' not ``first $z-1$''}
  virtual index of both with singlet states. This operator, which we call
  $\mathbb{E}_z$ and which matches the superoperator $\hat{\mathbb{E}}$ defined
  in \cite[, eq. 1.3]{fannes1992cayley} is a superoperator that acts on operators on
  $z-1$ qubits (by contraction with the first $z-1$ virtual indices of the
  projectors) and produces an operator on a single qubit.
  This produces $\tilde{\rho}_{m+1}\equiv \mathbb{E}_z(\rho_m\otimes\cdots\otimes\rho_m)$.
  \item We apply the 1D AKLT transfer matrix $\mathbb{E}_2$ $n$ times to 
  $\rho_m$ (incorporating a line of decorations), yielding the next-level
  boundary state $\rho_{m+1}\equiv {\mathbb{E}_2}^n\tilde{\rho}_{m+1}$.
  \item To evaluate the expectation value of an operator $O$ acting on the
  central site of the radius-$M$ system, we contract $O$ with the physical
  indices of the spin-$z/2$ projector and its conjugate and then contract
  the virtual indices with $z$ copies of $\rho_M$. (This yields
  $\bra{\Psi_M}O\ket{\Psi_M}$ for the non-normalized state $\Psi_M$;
  to normalize we perform the same contraction, but ommitting $O$ and
  directly contracting the physical indices of the projectors, to
  obtain $\braket{\Psi_M}{\Psi_M}$.)
\end{enumerate}
This procedure is a somewhat generalized version of that outlined in
Fig.~\ref{fig:branch-factor}.

\todo[inline]{Create tensor-network diagrams}

\begin{comment}
, is shown in Fig.~\ref{fig:TN-calculate}.
\begin{figure}
\missingfigure[figwidth=.3\textwidth]{contraction procedure}
\caption{Directly calculating the norm of the AKLT state}
%\todo{full description including $Y$,$Z$ vector}
\label{fig:TN-calculate}
\end{figure}
\end{comment}

For confirmation of the recurrence relation \eqref{eq:recurnz}, we will
calculate values of $m_z$ using \eqref{eq:recurnz} and \eqref{eq:sz-exact} and
plot those on top of the tensor-network calculations described above. (Since
both calculations are exact, the lines will not be distinguishable). When
$z\gtrsim 20$, these na\"ive tensor-network calculations are no longer
practical, and we will instead only use direct calculation from
\eqref{eq:recurnz}.

\begin{figure}
  \subfigure[$n=0$]{\includegraphics[width=.4\textwidth]{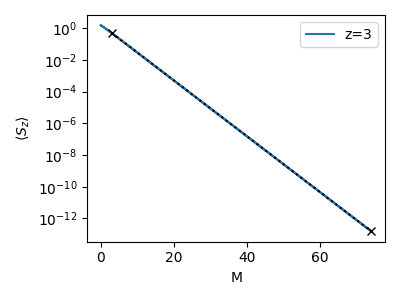}}
  \subfigure[$n=1$]{\includegraphics[width=.4\textwidth]{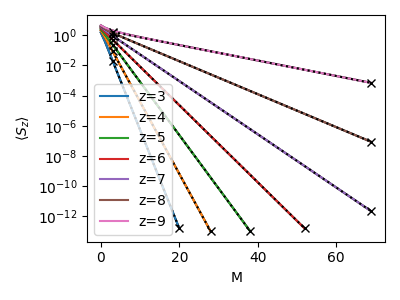}}
  \subfigure[$n=2$]{\includegraphics[width=.4\textwidth]{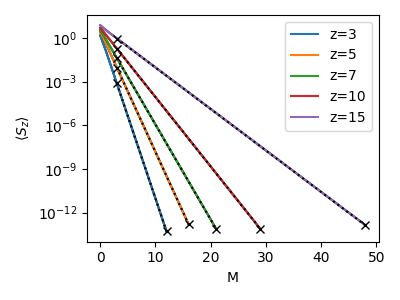}}
  \subfigure[$n=3$]{\includegraphics[width=.4\textwidth]{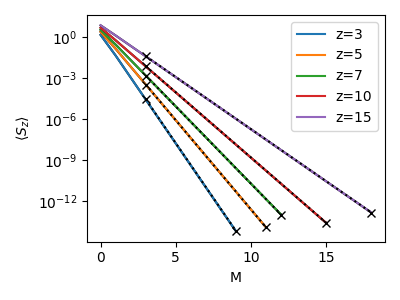}}
  \caption{Decay of the N\'eel order parameter for various numbers of
  decorations, within the disordered phase, calculated with tensor-network
  methods. The asymptotic 
  behavior dictated by \eqref{eq:mzdecay} is plotted with dotted lines and
  $\times$ markers.}
  \label{fig:decay-exp}
  \todo[inline]{Fix size, make asymptotic lines more visible}
\end{figure}

\begin{figure}
  \subfigure[$n=0$]{\includegraphics[width=.4\textwidth]{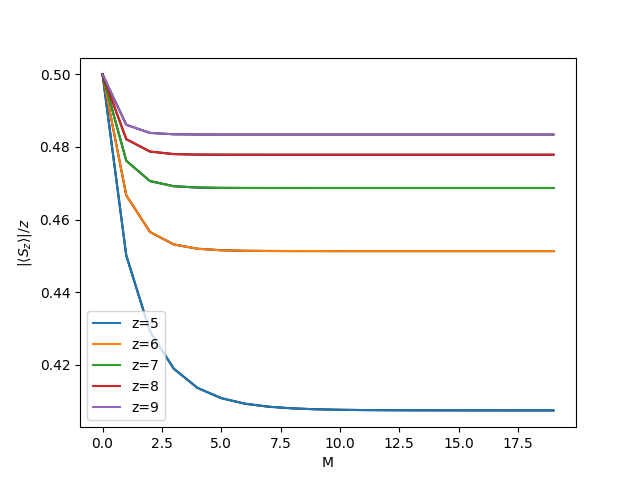}}
  \subfigure[$n=1$]{\includegraphics[width=.4\textwidth]{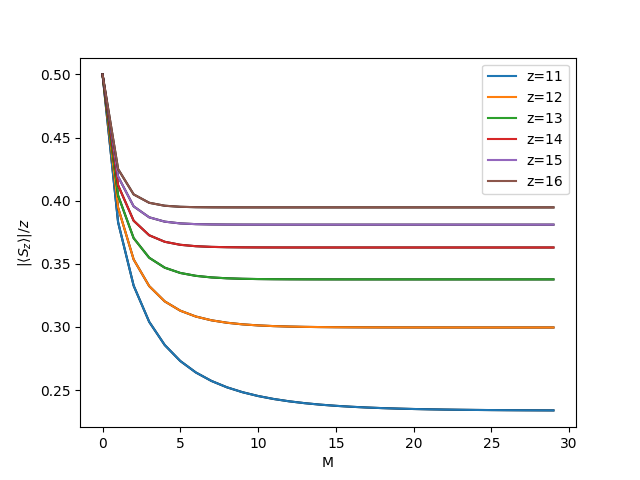}}
  \subfigure[$n=2$]{\includegraphics[width=.4\textwidth]{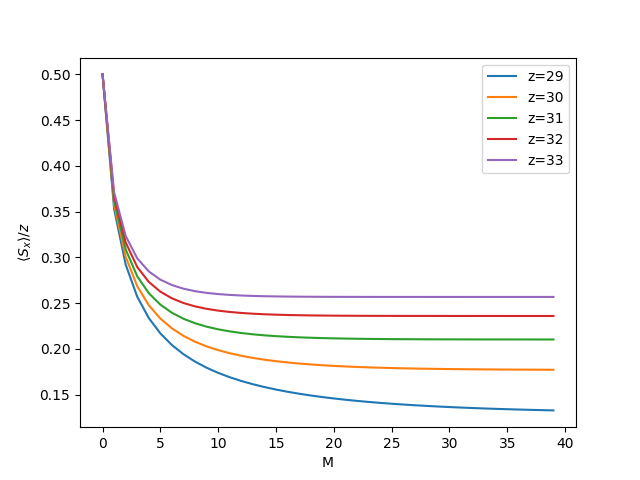}}
  \subfigure[$n=3$]{\includegraphics[width=.4\textwidth]{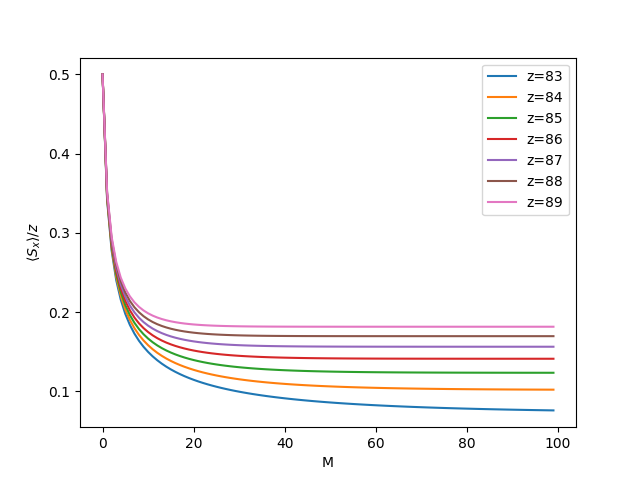}}
  \caption{Behavior of the N\'eel order parameter in the ordered phase.
  For (a) and (b) we use direct tensor-network calculations; for (c) and
  (d), order does not occur in cases where tensor-network computations
  are practical and so we use repeated application of the function
  $f_{n,z}$ instead.}
  \label{fig:ordered}
\end{figure}

\begin{figure}
  \includegraphics[width=.6\textwidth]{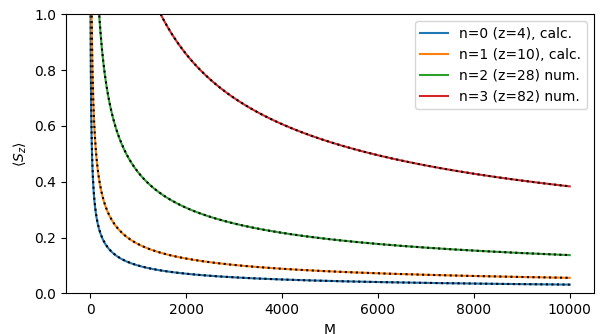}
  \caption{Behavior of the N\'eel order parameter at the phase boundary,
  with computed values compared with the theoretical asymptotic behavior
  predicted in \eqref{eq:mz-power-law} (dotted).
  When this cannot be probed with na\"ive tensor-network
  calculations ($n>1$), computation using \eqref{eq:fcont} is used instead.}
  \label{fig:decay-power}
\end{figure}

In Fig.~\ref{fig:decay-exp} we compare the large-$M$ behavior of the
N\'eel order parameter to that predicted by \eqref{eq:mzdecay}.
In Fig.~\ref{fig:ordered}, we confirm the prediction
that the state is antiferromagnetically ordered when $z > 3^{n+1}+1$:
in particular, the N\'eel order parameter converges to a nonzero expectation
value. In Fig.~\ref{fig:decay-power}, we confirm
the power-law relation of \eqref{eq:mz-power-law} for the critical case
$z=3^{n+1}+1$ for several values of $z$.

\begin{comment}
\begin{figure}
  \subfigure[$n=0,z=4$]{\includegraphics[width=.4\textwidth]{critdecayn0}}
  \subfigure[$n=1,z=10$]{\includegraphics[width=.4\textwidth]{critdecayn1}}
  \caption{Behavior of the N\'eel order parameter at the phase boundary,
  with computed values compared with the theoretical asymptotic behavior
  predicted in \eqref{eq:mz-power-law}.}
  \label{fig:decay-power}
\end{figure}

\begin{figure}
  \includegraphics[width=.7\textwidth]{critdecay_sim}
  \caption{Behavior of the N\'eel order parameter at the phase boundary
  for $n>2$, when this cannot be probed with na\"ive tensor-network
  calculations; computation using $f_{n,z}$ is used instead. Black dotted
  lines representing the asymptotic behavior in \eqref{eq:mz-power-law} are
  overlaid.}
  \label{fig:decay-power-sim}
  \todo[inline]{Combine with above into single plot? Omit ``asymp.'' labels?}
\end{figure}
\end{comment}

\clearpage
\section{The infinite state and the loop operator}
\label{sec:loop}
In the predictions of \eqref{eq:mzdecay} and \eqref{eq:mz-power-law}, we have
made use of the finite-size ``N\'eel order parameter'' of \cite{AKLT_1988} to,
for example, diagnose a distinction between ``critical'' behavior with
infinite ``correlation length'' and a more conventional disordered AKLT
phase with finite ``correlation length''. One may pose a reasonable objection
to this distinction that it is incompatible with taking the thermodynamic
limit: that is, the behavior in question is seen across length scales equal
to the system size, and so the possibility that it is due to finite-size
effects has not been excluded. We will therefore briefly introduce
a means of diagnosing these distinctions using the infinite state, which
\todo{as in the Gibbs or QMC formalism? add citations} can be defined for any
fixed-point boundary condition.

To summarize how this state is defined
\todo{rectify with sources + unify}, we use the superoperators
$\mathbb{E}_2$ and $\mathbb{E}_z$ defined above, with which we can
define the constraint on a pair of steady-state boundary density matrices
$\rho_\infty^+$ as
\begin{equation}
\begin{split}
\rho_\infty^+ &\propto \mathbb{E}_2^n\circ \mathbb{E}_z(\rho_\infty^-\otimes\cdots\otimes\rho_\infty^-)\\
\rho_\infty^- &\propto \mathbb{E}_2^n\circ \mathbb{E}_z(\rho_\infty^+\otimes\cdots\otimes\rho_\infty^+)
\end{split}
\label{eq:rho-steady}
\end{equation}
(We require a pair of density matrices due to the antiferromagnetic nature of
the AKLT state; when the system is disordered, or when $n$ is odd and thus the
spin-1 decorations mediate an effectively ferromagnetic interaction between
the ``main'' spin-$z/2$ sites, we expect
$\rho^+_\infty=\rho^-_\infty$.)

In particular, we may conclude that, for $W_\infty$ a fixed point of the
recurrence relation \eqref{eq:recurnz}, 
\begin{equation}
\rho^\pm_\infty \equiv \rho_z(W^{\pm 1}) \equiv \frac{1}{W_\infty^{1/2}+W_\infty^{-1/2}}\left(\begin{array}{ccc}W_\infty^{\mp\frac{1}{2}}&&0\\0&&W_\infty^{\pm\frac{1}{2}}\end{array}\right)
\label{eq:rho-inf}
\end{equation}
satisfies \eqref{eq:rho-steady}, as a consequence of our derivation of
\eqref{eq:recurnz}. (That is, they are defined by a ratio of eigenvalues
$W_\infty$.) For all values of $z$ and $n$, this means we can define a
disordered steady state $\rho_0=\frac{1}{2}\id$ from the trivial fixed point
$W=1$. When there is an additional fixed point $0\leq W_A<1$ representing 
antiferromagnetic order, this gives us an additional boundary steady state
-- or rather, applying $SU(2)$ transformations, an infinite family thereof.

Having established a boundary steady state\todo{note on uniqueness somewhere?},
we can define a reduced density matrix $\mathfrak{A}_M$\todo{think about
symbols} on a disc of radius $M$ as follows: we simply take the AKLT
construction on this disc and its conjugate and contract their free indices
pairwise with $\rho^s_\infty$ (where $s$ is the sign $(-1)^M$), and then
normalize. Then the expectation value $\langle O\rangle_\infty$ of an operator
$O$ with finite support can be defined by taking the follows. By definition,
there is some $M$ such that the support of $O$ is contained within the disc of
radius $M$. Thus, we state that
\[ \langle O\rangle_\infty=\op{tr}(\mathfrak{A}_M O). \]
We note that the construction of $\mathfrak{A}_M$ can be applied to boundary
density matrices which are not steady states, and define a (reduced) density
matrix $\mathfrak{B}_M(\rho)$, such that
\[ \mathfrak{B}_M(\rho^{(-1)^M}_\infty) = \mathfrak{A}_M. \]
This is similar to the construction used above to explicitly calculate the
N\'eel order parameter; in particular,\todo{Check sign + fencepost}
\[ m_z(W_M) = (-1)^M\op{tr}\left(\mathfrak{B}_M\left(\ket{\!\uparrow}\bra{\uparrow\!}\right)S_z^{(0)}\right) = (-1)^{M-N}\op{tr}\left(\mathfrak{B}_{M-N}(\rho_z(W_N))S_z^{(0)}\right). \]

With the infinite state defined, we will introduce the operator we will use to
diagnose long-range behavior. For a more physical setting than the Bethe
lattice, we might use a simple two-point correlator; however, as noted in
\cite{AKLT_1988}\todo{Refer to precise comments} these will display
exponential decay regardless of order. Instead, we simulate the kind of boundary
conditions we have been using to diagnose decay with a loop operator defined
as
\begin{equation}
L_M(\vec{r}) = \prod_{p\text{ at radius }M}\exp((-1)^{(n+1)M}\vec{r}\cdot\vec{S}^{(p)}),
\end{equation}
where $\vec{r}$ is some three-element vector and $\vec{S}^{(p)}$ is the
operator-valued vector $S_x\hat{x}^{(p)}+S_y\hat{y}^{(p)}+S_z\hat{z}^{(p)}$
acting on some (degree-$z$) site $p$. (The sign is inserted for consistency).
To analyze the behavior of the system when such an operator is applied,
we first see how the expressions in \eqref{eq:decoratedYZ} are modified when the
physical index is contracted with $\exp(-(-1)^{(n+1)M}\tau S_z)$:\todo{sign?}
\begin{equation}
\begin{split}
Y^{(0)}_1 &= (z-1)!\sum\limits_{m=0}^{z-1}(m+1)e^{-\tau (m+1-z/2)}(Y^{(n)}_{0})^m(Z_{0}^{(n)})^{z-m-1}\\
&= (z-1)!e^{-\tau (z/2+1)}(Z_{0}^{(n)})^z\sum\limits_{m=0}^{z-1}(m+1)(e^{-\tau}W_{0}^{(n)})^m \\
Z^{(0)}_1 &= (z-1)!\sum\limits_{m=0}^{z-1}(z-m)e^{-\tau (m-z/2)}(Y^{(n)}_{0})^m(Z_{0}^{(n)})^{z-m-1}\\
&= (z-1)!e^{-\tau z/2}(Z_{0}^{(n)})^z\sum\limits_{m=0}^{z-1}(z-m)(e^{-\tau}W_{0}^{(n)})^m.
\end{split}
\label{eq:YZwithloop}
\end{equation}
In other words, the relationship between $W^{(n)}_0$ and $W^{(0)}_1$, which
normally is given by the ``bare-lattice'' recurrence relation
$W^{(0)}_1=f_{0,z}(W^{(n)}_1)$, is instead given by $W^{(0)}_1=e^{-\tau}f_{0,z}(e^{-\tau}W^{(n)}_1)$. We can therefore capture the effect of the loop
operator on the parameter $W$ with
\begin{equation}
W \mapsto \tilde{f}_{n,z}(W,\tau) = {f_{0,2}}^n(e^{-\tau}f_{0,z}(e^{-\tau W})).
\end{equation}
This does not give us immediate information on expectation values
involving $L_M(\tau\hat{z})$, since the process of deriving those recurrence
relations involves implicit normalization. 
Nonetheless the construction of the density matrix
$\op{tr}_M(L_M(\tau\hat{z})\mathcal{A}_M)$ (where we have performed a
partial trace over the physical indices on sites at radius $M$ as well as those
on the decorations on segments connecting to those sites), once normalized,
proceeds identically to the construction of
$\mathfrak{B}_{M-1}(\tilde{\rho})$, with
$\tilde{\rho}\equiv\rho_z(\tilde{f}_{n,z}(W_\infty,\tau))$. We
therefore conclude that, for an operator $O$ supported within radius $M$,
\begin{equation}
\langle L_M(\tau \hat{z})O\rangle_\infty = \op{tr}(\mathfrak{B}_M(\tilde{\rho})O)\langle{L_M(\tau\hat{z})}\rangle_\infty.
\end{equation}
In particular, while the behavior of $\langle L_M(\tau\hat{z}) \rangle$, and
therefore of $\langle L_M(\tau\hat{z})O\rangle$, will typically be dominated by
short-range effects due to the exponential scaling $z(z-1)^{M-1}$ of the number of
sites at radius $M$ in the Bethe lattice, we can effectively isolate long-range
effects of the loop operator with the functional
\begin{equation}
L_{M,\vec{r}}[O] \equiv \frac{\langle L_M(\vec{r})O\rangle_\infty}{\langle L_M(\vec{r})\rangle_\infty} = \op{tr}(\mathfrak{B}_M(\tilde{\rho})O)
\label{eq:loop-op-def}
\end{equation}
(with the latter equality in the case of $\vec{r}=\tau\hat{z}$).
We will call this the \textit{loop expectation value}.
As an example, we will use this to demonstrate critical decay in
Fig.~\ref{fig:crit-continuous}.

\todo[inline]{Reproduce earlier plots with loop expectation value}

As a final note, we may use these loop operators to more precisely infer the
behavior of two-point functions. In the small-$\tau$ limit, we may
generally represent $\tilde{f}_{n,z}(W_\infty,\tau)$ as a small correction
with first-order behavior $W_\infty - \alpha\tau$. Since we already know
the leading-order behavior of $f_{n,z}$ and $m_z$ in the disordered case
$W_\infty=1$, we may conclude
from \eqref{eq:S-order1} and \eqref{eq:mzdecay} that
\[ L_{M,\tau \hat{z}}[S_z^{(0)}] \simeq \alpha\tau \frac{z(z+1)}{3}\left(\frac{z-1}{3^{n+1}}\right)^M \sim \tau e^{-M/\xi}. \]
(We include the cases where $W=1$ is a marginal or unstable fixed point, in
which $\xi=\infty$ and $\xi<0$ respectively.)
In the same limit, the loop operator, being a product of exponentials in
$\tau$, is straightforward to expand to first order:
\[ L_{M}(\tau\hat{z}) = 1 - (-1)^{(n+1)M}\tau \sum_{p\text{ at radius }M}S_z^{(p)} + \mathcal{O}(\tau^2). \]
Since, again confined to the disordered state where $W=1$, one-point functions
vanish, this implies that the loop expectation value behaves as
\[ L_{M,\tau\hat{z}}[S_z^{(0)}] = 1 - (-1)^{(n+1)M}\tau\sum_{p\text{ at radius }M}\langle S_z^{(0)}S_z^{(p)}\rangle_\infty. \]
Equating these two approximations, and noting that the symmetry of the Bethe
lattice implies that the above two-point function will be identical for any
$p$, we find
\[ \lim_{\tau\to 0}\tau^{-1}L_{M,\tau\hat{z}}[S_z^{(0)}] = -(-1)^{(n+1)M}z(z-1)^{M-1}\langle S_z^{(0)}S_z^{(p)}\rangle_\infty = \alpha\frac{z(z+1)}{3}e^{-M/\xi}, \]
and therefore,\todo{Need to work on those signs}
\begin{equation}
\langle S_z^{(0)}S_z^{(p)} \rangle_\infty = -(-1)^{(n+1)M}\alpha\frac{z^2-1}{3}e^{-(1/\xi+\log(z-1))M}.
\end{equation}
\todo[inline]{Plot verification}
\todo[inline]{Is this exact?}
Note that the actual form of the effective correlation length
\[\xi^{\prime-1}=\xi^{-1}+\log(z-1)=(n+1)\log3 \]
provides a re-derivation of the trivial prediction found in \cite{AKLT_1988},
i.e. a two-point function that (in the undecorated case) decays as
$3^{-M}$. However, we can still use this to conclude that, for Bethe lattice
systems of degree $z$ more generally,
with $O$ some order parameter of spontaneously-broken
symmetry, if
\begin{equation}
\langle O^{(0)} O^{(p)} \rangle \sim e^{-M/\xi'},\ 
\left\{\begin{array}{ccl}
\xi^{\prime-1}>\log(z-1)&\Rightarrow&\text{disorder},\\
\xi^{\prime-1}=\log(z-1)&\Rightarrow&\text{criticality, and}\\
\xi^{\prime-1}<\log(z-1)&\Rightarrow&\text{order}.\\
\end{array}\right.
\end{equation}
\todo[inline]{Is this at all noteworthy?}

Additionally, when the system is in an ordered configuration with
polarization $m_0 = m_z(W_\infty)$, we can expect -- again for $M$ large but
fixed --\todo{Clean up this derivation a bit. Also maybe the other ones.}
\begin{align}
%m_z(W_M) &\simeq m_0 + m_1e^{-M/\xi_A}\notag\\
m_z(f_{n,z}^M(W_\infty + \alpha \tau)) &\simeq m_0 + \alpha\tau e^{-M/\xi_A}\notag\\
\langle L_M(\tau\hat{z})S_z^{(0)}\rangle_\infty &\simeq \langle S_z^{(0)}\rangle_\infty - (-1)^{n(M+1)}z(z-1)^{M-1}\tau\langle S^{(0)}_zS_z^{(p)}\rangle_\infty\notag\\
\langle L_M(\tau\hat{z})\rangle_\infty &\simeq 1 - (-1)^{n(M+1)}z(z-1)^{M-1}\tau\langle S^{(p)}_z\rangle_\infty\notag\\
L_{M,\tau\hat{z}}[S_z^{(0)}] &\simeq m_0 - (-1)^{n(M+1)}z(z-1)^{M-1}\tau(\langle S^{(0)}_zS^{(p)}_z\rangle_\infty - m_0^2)\notag\\
\langle S^{(0)}_zS^{(p)}_z\rangle_\infty &\simeq m_0^2 - \frac{z-1}{z}\alpha e^{-M(1/\xi_A+\log(z-1))},
\end{align}
where the effective correlation length $\xi_A$, though in practice difficult
to calculate analytically, is defined by $\xi_A^{-1}=1-f'_{n,z}(W_\infty)$.
\section{Interpolating $n$}
\label{sec:interpolating}
We have, in \eqref{eq:fprime}, an appealingly simple criterion for the
presence of order: a line $3^{n+1} = z-1$ separating order and disorder
in all cases. Indeed, we find that nearly all the expressions we have used
to determine the behavior of these decorated systems have a rational
dependence on $3^n$. Therefore, if we could somehow make the parameter
$\Gamma=3^n$ continuous, we would expect a phase transition for
any $z>4$ at the critical point
\[ \Gamma = \frac{z-1}{3}. \]
Instead of using this parameter $\Gamma$, we will find it more practical to 
employ a parameter\footnote{There are various reasons for using $\gamma$ over
the more obvious $\Gamma$; it will simplify some equations and make the
``antiferromagnetic-ferromagnetic'' duality that will arise much simplier. For
now we will simply say that the ``bare'' case $\gamma=0$ ($\Gamma=1$) is a
much firmer boundary than the ``decoupling'' point $\gamma=1$
($\Gamma\to\infty$).}
\begin{equation}
\begin{split}
\gamma &\equiv \frac{\Gamma-1}{\Gamma+1} = \frac{3^n-1}{3^n+1}\text{, or}\\
3^n &= \frac{1+\gamma}{1-\gamma}.
\end{split}
\label{eq:gamma-def}
\end{equation}
This allows us to rewrite \eqref{eq:recurnz} as
\begin{equation}
f_{\gamma,z}(W) = \frac{\sum\limits_{i=0}^{z-1}\left[(z-i)\gamma+(i+1)\right]W^i}{\sum\limits_{i=0}^{z-1}\left[(i+1)\gamma+(z-i)\right]W^i}.
\label{eq:fcont}
\end{equation}
\todo[inline]{Specify conditions for using $f_n$ vs $f_\gamma$ or use a
different symbol (?)}
\todo[inline]{Ignoring physics on decoration}
The two methods we use to reproduce this, in part or in whole,
are detailed below.

\subsection{A deformed decoration}
\label{sec:deformed}
We may seek to emulate \eqref{eq:transfer} by directly modifying
\eqref{eq:decoratedYZ}. The coefficients applied to $(Y^{(k-1)},Z^{(k-1)})$
there come directly from elements of the spin-1 projector. Modifications made
to this object with an onsite deformation much like those studied in
\cite{decorated1,decorated2,guo2020aklt}.
In particular, a diagonal (in the $z$ basis
deformation will alter \eqref{eq:recur-bare} by multiplication on the
appropriate term (by the square of element in question). Thus, with a single
decorated, deformed with a matrix defined by (real) diagonal elements
$(1,a,1)$, \eqref{eq:transfer} becomes
\todo[inline]{Explain this more clearly}
\[
\left(\begin{array}{c} Y_M^{(1)}\\ Z_M^{(1)} \end{array}\!\right) 
=
\left(\begin{array}{ccc} 2&&a^2 \\ a^2&&2 \end{array}\right)
\left(\begin{array}{c} Y_M^{(0)}\\ Z_M^{(0)} \end{array}\!\right).
\]
Referring again to \eqref{eq:transfer}, the desired matrix in this expression
would be proportional to 
\begin{equation}
\left(\begin{array}{cc} \frac{3^n+1}{2}&\frac{3^n-1}{2} \\ \frac{3^n-1}{2}&\frac{3^n+1}{2} \end{array}\right)
= \frac{3^n+1}{4}\left(\begin{array}{cc}2&2\gamma\\2\gamma&2\end{array}\right),
\end{equation}
which is to say that we can achieve the behavior in \eqref{eq:fcont} by
deforming the $m_z=0$ index of a single decoration by $a=\sqrt{2\gamma}$.

\subsection{Interpolating a single decoration}
\label{sec:frustrated}
A deformation like the one used above explicitly breaks the $SU(2)$ down to
$O(2)$. We will explore the consequences of doing so later\todo{reference
section?}, but for now we will seek a way of replicating this behavior
that is $SU(2)$-invariant.

In order to do so, we analyze the (full) transfer matrices
on the decorated edge. In a basis
$(\ket{\up}\bra{\up}, \ket{\up}\bra{\down},
\ket{\down}\bra{\up}, \ket{\down}\bra{\down})$\todo{fix this}
for the boundary density matrices,
we can write the transfer matrix of a single
decoration as $s_2A_2s_2$, where
\[ s_2 = 
\left(\begin{array}{cccc}
0&0&0&1\\
0&0&-1&0\\
0&-1&0&0\\
1&0&0&0
\end{array}\right)
\]
is the tensor product of two singlet states and
\[
A =
\left(\begin{array}{cccc}
1&0&0&\frac{1}{2}\\
0&0&\frac{1}{2}&0\\
0&\frac{1}{2}&0&0\\
\frac{1}{2}&0&0&0\end{array}\right)
\]
is the result of contracting two spin-1 projectors.

In this basis, $s_2$ and $A_2$ commute,\footnote{We note that we must use
caution when treating $s_2$ and $A_2$ as endomorphisms or square matrices;
rather than both belonging to $\op{Hom}(V,V)\cong V\otimes V^*$, they belong
to $V\otimes V$ and $V^*\otimes V^*$, respectively, with
$V$ being the $\sfrac{1}{2}\otimes\sfrac{1}{2}^*$ representation of $SU(2)$.}
and so the transfer matrix for $n$ decorations may be written
\[ (s_2A_2)^ns_2 = A_2^ns_2^{n+1}. \]
Blockwise analysis of $A_2$ and $s_2$ leads us to conclude
\begin{align}
A_2^n &=
\frac{1}{2^{n+1}}\left(\begin{array}{cccc}
3^n+1 &0&0& 3^n-1\\
0& 1+(-1)^n & 1-(-1)^n &0\\
0& 1-(-1)^n & 1+(-1)^n &0\\
3^n-1 &0&0& 3^n+1
\end{array}\right)\\
A_2^ns_2^{n+1} &=
\frac{1}{2^{n+1}}\left(\begin{array}{cccc}
3^n-(-1)^n &0&0& 3^n+(-1)^n\\
0& 0 & 2(-1)^{n+1} &0\\
0& 2(-1)^{n+1} & 0 &0\\
3^n+(-1)^n &0&0& 3^n-(-1)^n
\end{array}\right)
\end{align}
Crucially, these transfer matrices span a two-dimensional real space. We should
therefore be able to obtain those by combining the $n=0$ and $n=1$ cases, using
a ``perturbation'' parameter $\delta^2$ and an unstated normalization factor:
\begin{equation}
s_2+\delta^2s_2A_2s_2 =
\left(\begin{array}{cccc}
\delta^2&0&0&\frac{\delta^2}{2}+1\\
0&0&\frac{\delta^2}{2}-1&0\\
0&\frac{\delta^2}{2}-1&0&0\\
\frac{\delta^2}{2}+1&0&0&\delta^2
\end{array}\right)
\propto
\left\{
\begin{array}{ccl}
\frac{3^n+1}{2^{n+1}}
\left(\begin{array}{cccc}
\gamma&0&0&1\\
0&0&\gamma-1&0\\
0&\gamma-1&0&0\\
1&0&0&\gamma
\end{array}\right)&,&n$ even$\\[2.5em]
\frac{3^n+1}{2^{n+1}}
\left(\begin{array}{cccc}
1&0&0&\gamma\\
0&0&1-\gamma&0\\
0&1-\gamma&0&0\\
\gamma&0&0&1
\end{array}\right)&,&n$ odd$
\end{array}\right.
\end{equation}
Combining a single decoration with the undecorated edge has the effect of 
frustrating the antiferromagnetic AKLT interactions; in particular we expect
an antiferromagnetic system for $\delta \ll 1$ and a ferromagnetic system for
$\delta \gg 1$. We can then establish two correspondences:
\begin{equation}
\delta^2 =
\left\{\begin{array}{ccr}
\frac{2\gamma}{2-\gamma}&,&n$ even$\\[1ex]
\frac{2}{2\gamma-1}&,&n$ odd$
\end{array}\right..
\label{eq:interpolation-values}
\end{equation}
\subsubsection{The ferromagnetic-antiferromagnetic ``duality''}
It is of some interest that the two expressions in
\eqref{eq:interpolation-values} are related to one another by the exchange
$\gamma\leftrightarrow\gamma^{-1}$. Looking to \eqref{eq:fcont}, we realize
that
\[ f_{\gamma^{-1},z}(W) = f_{\gamma,z}(W^{-1}) = f_{\gamma,z}(W)^{-1}. \]
That is, this relation interchanges an alternating polarization with a
monotonic one. (Since $f$ is based on the AKLT-norm formula
\eqref{eq:norm-loops} which flips the spin orientation on alternating sites,
the former actually represents ferromagnetic interactions and the latter
antiferromagnetic, at least for $n=0$. Since the unperturbed case in the
frustration method of realizing continuous $\gamma$ is $n=0$, whereas
the unperturbed case in the deformation method is $n=1$, we will
consider the antiferromagnetic regime to be $\gamma<1$ in the former
case and $\gamma>1$ in the latter, a choice which could not possibly lead
to any confusion.)

This is related to the fact that the derivation of this transfer matrix, and
the various other expressions related to $z$-polarization we have used,
effectively only rely on the elements acting on $\{\ket{\up}\bra{\up},
\ket{\down}\bra{\down}\}$ (or, in terms of \eqref{eq:norm-loops}, cases where
$G=G'$). When we are considering observables diagonal in the $z$ basis acting
on states that are disordered or $\pm z$-polarized, this means that expectation
values can be exactly mapped onto one another under this ``duality'', after
flipping spin orientations on alternating sites. Due to $SU(2)$ invariance, we
can do this with states and observables of any single polarization. Simple
arguments will show that a few additional observables, such as
$\langle \vec{r}\cdot \vec{S}\rangle$, can be related in this way (as the
contribution from the component of $\vec{r}$ perpendicular to the direction of
polarization will vanish), but there is no more general duality.
\todo{Check, maybe plot?}
Nonetheless, we will make our default choice of correspondence
\[ \delta^2 = \frac{2\gamma}{2-\gamma}, \]
such that $\gamma<1$ is antiferromagnetic and $\gamma>1$ ferromagnetic,
and such that the correspondence with the original $n$-times decorated system
is incomplete when $n$ is odd.

We further take note of the special ``self-dual'' case $\gamma=1$, corresponding
to the $n\to\infty$ limit. Due to the finite correlation length of the AKLT
chain, we would expect this to act to decouple otherwise-neighboring
Bethe lattice spins, and we can demonstrate that this is the case. Generally,
we can treat the decoration as as a mapping from a pair of virtual spins to a
singlet-triplet space $\op{span}\{\wket{s},\wket{-},\wket{0},\wket{+}\}$,
equal to
\[ \wket{0}\left(\bra{\up\down}-\bra{\down\up}\right) + \delta\wket{+}\bra{\down\down} + \delta\wket{-}\bra{\up\up} - \delta\sqrt{\sfrac{1}{2}}\wket{0}\left(\bra{\up\down} + \bra{\down\up}\right). \]
Then, when $\gamma=1$ and so $\delta=\sqrt{2}$, this is proportional to a
unitary transformation. Therefore, this effectively decouples the order-$z$
spins from each other.\todo{Is this completely true? Are they more interesting
gauge dof in some way?}

\todo[inline]{Is there some form of gauge invariance here in the deformed realization?}

\todo[inline]{Hamiltonian -- is it always sufficient? including in self-dual case?}

\subsection{The phase transition}
As alluded to above\todo{connect better}, realizing \eqref{eq:fcont} yields a
transition at $\Gamma_c=\sfrac{(z-1)}{3}$, or
\[ \gamma_c = \frac{z-4}{z+2}. \]
(We note that this also implies, by the ferromagnetic-antiferromagnetic
``duality'' we have just introduced, another transition at
$\gamma=\sfrac{(z+2)}{(z-4)}$.)

\begin{figure}
  \subfigure[$z<10$]{\includegraphics[width=.4\textwidth]{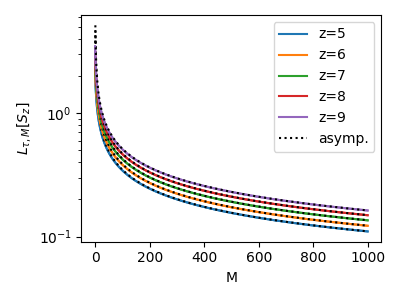}}
  \subfigure[$z>10$]{\includegraphics[width=.4\textwidth]{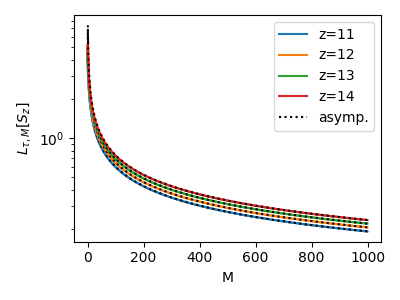}}
  \caption{Behavior of the ``order parameter'' \eqref{eq:loop-op-def},
  with $\tau=1$, at the critical point for various $z$
  in cases not coinciding with integer $n$. Computations using
  both types of decoration are shown, but cannot be distinguished from
  each other; the asymptotic behavior in \eqref{eq:mz-power-law} is overlaid.}
\label{fig:crit-continuous}
\end{figure}

At this critical point we find that the leading-order behavior is still
governed by the power-law decay in \eqref{eq:mz-power-law}, as we confirm
in Fig.~\ref{fig:crit-continuous}.

Given the presence of a continuous parameter, we may determine some additional
critical exponents. First, in the disordered phase, i.e. when
$\gamma_c<\gamma<1$, we can rephrase \eqref{eq:mzdecay} as the presence of a
sort of correlation length: noting that the asymptotic behavior can be written
$m_z\sim (\sfrac{\Gamma_c}{\Gamma})^M$, we determine
\begin{equation}
\xi^{-1} = \log(1+\gamma)-\log(1-\gamma)-\log(1-\gamma_c)+\log(1+\gamma_c) \simeq \frac{2}{1-\gamma_c^2}(\gamma-\gamma_c)
\label{eq:xi-divergence}
\end{equation}
\begin{figure}
  \subfigure[$z=4$]{\includegraphics[width=.3\textwidth]{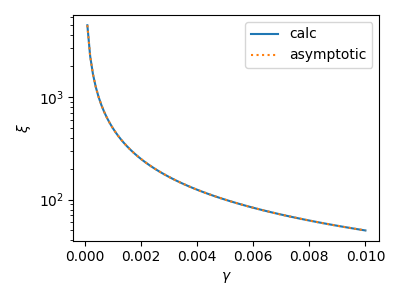}}
  \subfigure[$z=5$]{\includegraphics[width=.3\textwidth]{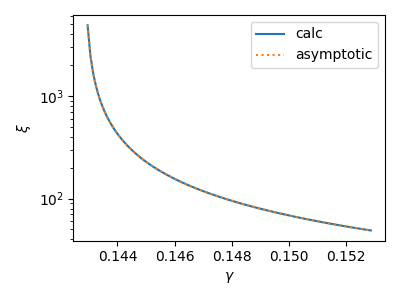}}
  \subfigure[$z=6$]{\includegraphics[width=.3\textwidth]{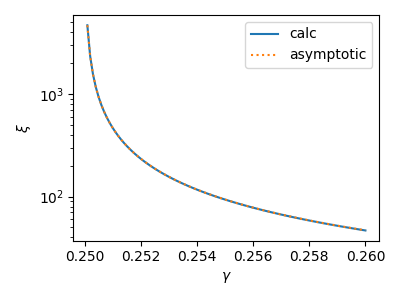}}
  \subfigure[$z=7$]{\includegraphics[width=.3\textwidth]{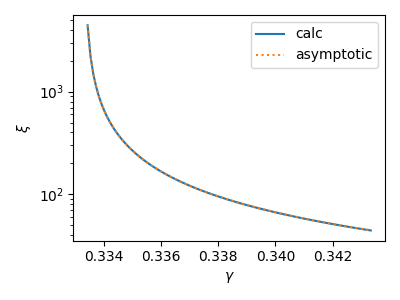}}
  \subfigure[$z=8$]{\includegraphics[width=.3\textwidth]{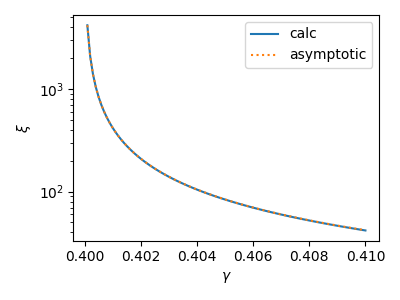}}
  \subfigure[$z=9$]{\includegraphics[width=.3\textwidth]{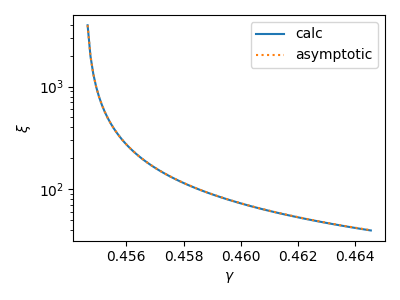}}
  \caption{The divergence of the correlation length $\xi$, with values
  computed directly from the recurrence relation \eqref{eq:fcont}, compared
  with the predicted asymptotic values (dotted) shown in \eqref{eq:xi-divergence}.}
  \label{fig:xi-diverge}
\end{figure}
in the vicinity of the critical point: that is, as shown in
\ref{fig:xi-diverge}, \todo{figure with
divergence of correlation length} the correlation length diverges as
$\xi\sim \Delta\gamma^{-1}$.

\begin{figure}
  \subfigure[$z=5$]{\includegraphics[width=.4\textwidth]{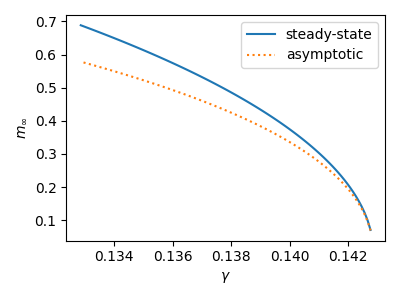}}
  \subfigure[$z=6$]{\includegraphics[width=.4\textwidth]{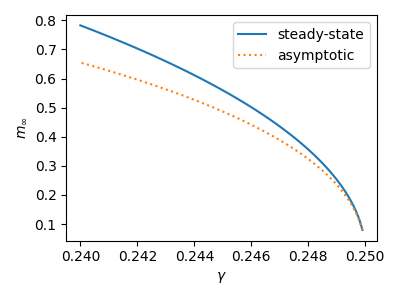}}
  \subfigure[$z=7$]{\includegraphics[width=.4\textwidth]{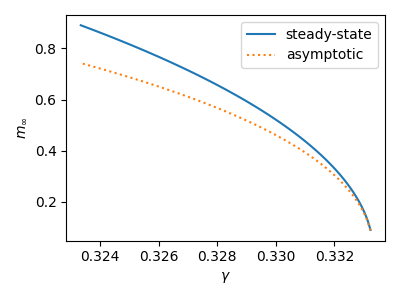}}
  \subfigure[$z=8$]{\includegraphics[width=.4\textwidth]{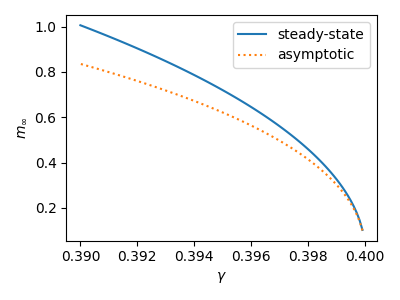}}
  \caption{Behavior of the steady-state magnetization ($\langle S_z\rangle$
  for the $z$-polarized state) in the ordered phase
  near the critical point, for various values of $z$. We plot results from
  both realizations, from direct calculation with $f_{\Gamma,z}$,
  and with the approximation \eqref{eq:orderedapprox}.}
  \label{fig:order-S-continuous}
\end{figure}

Meanwhile, we may now begin in earnest to study the behavior of the system
within the ordered phase ($\gamma < \gamma_c$), since we may be able to
determine the location of the attractive fixed point for $\gamma_c-\gamma$
sufficiently small - at least, so long as $\gamma_c$ is truly a second-order
transition, as the critical behavior suggests. We analyze the ordered fixed
point near the critical point using the third-order expansion
\eqref{eq:numorder3} of the numerator of $f_{\gamma,z}(W)-W$, with
$W=1-\varepsilon$ and $\Delta\gamma=\gamma-\gamma_c$ ($0<-\Delta\gamma\ll 1$),
replacing $3^{n+1}-z+1$ for the time with
$3\Delta\Gamma=\Delta\gamma\sfrac{(z+2)^2}{6}$. Then the fixed point is
approximately determined by $0<\varepsilon\ll 1$ such that
\[ z(z+1)\Delta \Gamma\;\varepsilon_c - z(z^2-1)\frac{\Delta\Gamma}{2}\varepsilon_c^2 + z(z^2-1)(z-2)\left[\frac{\Delta\Gamma}{6} + \frac{z+2}{180}\right]\varepsilon_c^3 = 0.\]

We solve this to get an approximation to the fixed point
$W_\infty \simeq 1-\varepsilon_c$, where
\begin{align}
\varepsilon_c &= \frac{45\Delta\Gamma + 6\sqrt{-5\frac{\Delta\Gamma}{z-1}}\sqrt{\frac{15}{4}(5z-13)\Delta\Gamma + z^2-4}}{(z-2)(30\Gamma + z+2)}
\label{eq:orderedapprox}\\
&\simeq 6(-\Delta \Gamma)^{\sfrac{1}{2}}\sqrt\frac{5}{(z-1)(z^2-4)}\simeq (\gamma_c-\gamma)^{\sfrac{1}{2}}\sqrt\frac{10(z+2)}{(z-1)(z-2)}.
\label{eq:orderedorder1}
\end{align}

We confirm this numerically in Fig.~\ref{fig:order-S-continuous}.

\section{Competing order in the anisotropic realization}
\label{sec:anisotropic}
\begin{comment}
\begin{figure}
    \subfigure{\includegraphics[width=.4\textwidth]{anisz8g2,6}}
    \subfigure{\includegraphics[width=.4\textwidth]{anisz8g3}}
    \subfigure{\includegraphics[width=.4\textwidth]{anisz8g3,6}}
    \subfigure{\includegraphics[width=.4\textwidth]{anisz8g4}}
    \caption{Decay of $\langle S_x\rangle_M$, in both the anisotropic and
    isotropic implementations of continuous $\Gamma$, at coordination number
    $z=8$, within the region where both are disordered.}
    \label{fig:decay-anisotropic}
\end{figure}
\end{comment}

\begin{figure}
    \subfigure[$z=8$]{\includegraphics[width=.3\textwidth]{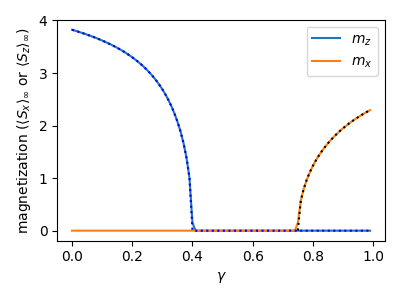}}
    \subfigure[$z=9$]{\includegraphics[width=.3\textwidth]{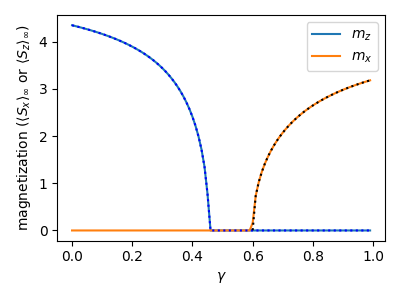}}
    \subfigure[$z=10$]{\includegraphics[width=.3\textwidth]{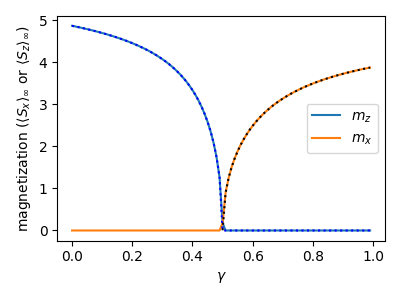}}
    \subfigure[$z=11$]{\includegraphics[width=.3\textwidth]{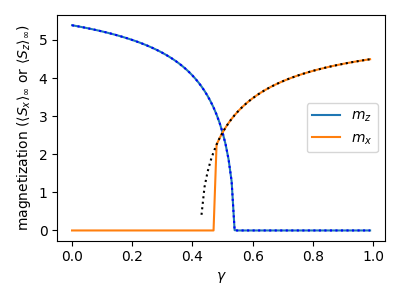}}
    \subfigure[$z=12$]{\includegraphics[width=.3\textwidth]{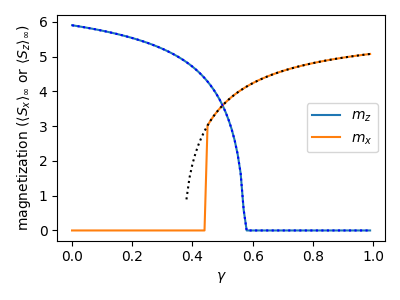}}
    \subfigure[$z=13$]{\includegraphics[width=.3\textwidth]{anisz12}}
    \caption{Limiting values of the $x$ and $z$ magnetization for
    coordination numbers 8-13, for the anisotropic implementation of continuous
    $\gamma$.In $z=11$ and
    $z=12$, when approaching the $x$ transition from the ordered side, we
    see an aberration due to spontaneous symmetry breaking, i.e. the
    system becomes polarized in the $z$ direction and therefore loses
    polarization in the $x$ direction.}
    \label{fig:phases-anisotropic}
\end{figure}
In subsection~\ref{sec:deformed}, we realized the recurrence relation
\eqref{eq:fcont} by deforming the AKLT state on the singly-decorated Bethe
lattice. Unlike the realization in subsection~\ref{sec:frustrated}, this
breaks the $SU(2)$ symmetry of the AKLT state down to the $O(2)$ subgroup
(generated by arbitrary rotations about the $z$ axis and the $\pi$ rotation
about, e.g., the $x$ axis.) The conclusions we obtained from this analysis
determine when we observe magnetic polarization of along the symmetry axis,
spontaneously breaking the $\mathbb{Z}_2$
quotient group; e.g., for coordination number $z>10$, when the model with
$n=1$ isotropic decoration (corresponding to parameter $\gamma=\sfrac{1}{2}$)
exhibits spontaneous breaking of $SU(2)$, we find that we can tune
$\gamma$ up to a critical point $\gamma_c$ above which the state is disordered,
whereas for the case $z=10$ where the isotropic model is critical, we find
that spontaneous symmetry breaking occurs at any value $\gamma<\sfrac{1}{2}$.
It is natural, then, to ask when polarization in the basal $xy$ plane occurs,
particularly given that we know it does in the
aforementioned isotropic $z>10$ cases.

We will primarily consider polarization along a single axis within the $xy$
plane: without loss of generality, the $x$ axis. When we wish to examine
the system in the presence of possible $x$ polarization, and/or under
the action of some combination of $S_x$ operators, we must restore the sum
over the second subset $G'$ in \eqref{eq:norm-loops}\todo{elaborate?}.
That is, when we factor out branches of the Bethe lattice as in \eqref{eq:Yfull}
and \eqref{eq:Zfull}, we must consider factors, which I will label $X^{\pm}_M$,
corresponding to the cases where $G\neq G'$ on the edge in queston:
\[ 
X_M^\pm = \sum_{G,G'}\delta_{m_0(G),m_0(G')\pm 1}c_z(m_0(G)+\tfrac{1\mp 1}{2})\prod_{k\in B_M}\delta_{m_k(G),m_k(G')}c_z(m_k(G)) \]
(where again $c_z(m)\equiv m!(z-m)!$ for convenience).
Ultimately this leads us to rewrite \eqref{eq:recur-bare},
cataloging 
the possible options\todo[disable]{Clean up the language} by numbers $m$ set to equal $m_0(G)$,
i.e. the total number of edges in $G$ incident on the root vertex, and
$i$ set to equal the number of \textit{pairs} of edges on which $G$ and
$G'$ do not match:
\begin{align*}
Y_{M+1} &= \sum_{i=0}^{\lfloor\sfrac{z-1}{2}\rfloor}\sum_{m=i}^{z-i-1}\binom{z-1}{m}\binom{m}{i}\binom{z-m-1}{i}(m+1)!(z-m-1)!Y_M^{m-i}Z_M^{z-m-i-1}(X_M^+)^i(X_M^-)^i\\
Z_{M+1} &= \sum_{i=0}^{\lfloor\sfrac{z-1}{2}\rfloor}\sum_{m=i}^{z-i-1}\binom{z-1}{m}\binom{m}{i}\binom{z-m-1}{i}m!(z-m)!Y_M^{m-i}Z_M^{z-m-i-1}(X_M^+)^i(X_M^-)^i\\
X^+_{M+1} &= \sum_{i=0}^{\lfloor \sfrac{z}{2}\rfloor-1}\sum_{m=i}^{z-i-2}\binom{z-1}{m}\binom{m}{i}\binom{z-m-1}{i+1}(m+1)!(z-m-1)!Y_M^{m-i}Z_M^{z-m-i-2}(X_M^+)^i(X_M^-)^{i+1}\\
X^-_{M+1} &= \sum_{i=0}^{\lfloor \sfrac{z}{2}\rfloor-1}\sum_{m=i+1}^{z-i-1}\binom{z-1}{m}\binom{m}{i+1}\binom{z-m-1}{i}m!(z-m)!Y_M^{m-i-1}Z_M^{z-m-i-1}(X_M^+)^{i+1}(X_M^-)^i.
\end{align*}

On the lattice with a single decoration deformed as above by
$a$, we extend the remaining two equations from \eqref{eq:decoratedYZ} by
setting $z=2$ and multiplying terms corresponding to $m_i=1$ by
$a^2=2\gamma$:
\begin{align*}
Y_M^{(1)}&=2Y_M^{(0)}+2\gamma Z_M^{(0)}\\
Z_M^{(1)}&=2\gamma Y_M^{(0)}+2Z_M^{(0)}\\
X_M^{\pm(1)}&=2\gamma X_M^{\mp(0)}
\end{align*}
We have seen that reflection-invariance in the $z$ direction is
equivalent to $Y=Z$; we may also see that reflection-invariance in the $y$
direction is equivalent to $X^+=X^-$, and indeed, the above
equations preserve these relations. 

Then we can turn the above recursive equations into
\begin{align*}
Y_{M+1}^{(0)} &= \sum_{i=0}^{\lfloor\sfrac{z-1}{2}\rfloor}(Y_M^{(1)})^{z-2i-1}(X_M^{(1)})^{2i}\sum_{m=i}^{z-i-1}\frac{(z-1)!(m+1)!(z-m-1)!}{i!^2(m-i)!(z-m-i-1)!}\\
X_{M+1}^{(0)} &= \sum_{i=0}^{\lfloor \sfrac{z}{2}\rfloor-1}(Y_M^{(1)})^{z-2i-2}(X_M^{(1)})^{2i+1}\sum_{m=i+1}^{z-i-1}\frac{(z-1)!m!(z-m)!}{i!(i+1)!(m-i-1)!(z-m-i-1)!}\\
Y_M^{(1)} &= 2(1+\gamma)Y_M^{(0)}\\
X_M^{(1)} &= 2\gamma X_M^{(0)},
\end{align*}
or, defining a single parameter $V_M=X_M^{(1)}/Y_M^{(1)}$,
\begin{align*}
Y^{(1)}_{M+1}
  &= 2(1+\gamma)(z-1)!(Y_M^{(1)})^{z-1}\sum_{i=0}^{\lfloor\sfrac{z-1}{2}\rfloor} (i+1) V_M^{2i}
     \sum_{n=i}^{z-i-1} \binom{n+1}{i+1} \binom{z-n-1}{i}\\
X^{(1)}_{M+1}
  &= 2\gamma(z-1)!(Y_M^{(1)})^{z-1}\sum_{i=0}^{\lfloor \sfrac{z}{2}\rfloor-1} (i+1) V_M^{2i+1}
    \sum_{n=i+1}^{z-i-1} \binom{n}{i+1} \binom{z-n}{i+1}\\
\end{align*}
Taking the quotient, and applying a combinatorial
identity, we obtain
\todo[inline]{explain, possibly derive?}
\begin{equation*}
V^{(1)}_{M+1} = \dfrac{(1+\gamma)\sum\limits_{j=0}^z \frac{1+(-1)^j}{2} \frac{j+2}{2} \binom{z+1}{j+2} (V_M^{(1)})^j}
  {\gamma\sum\limits_{j=0}^z \frac{1-(-1)^j}{2} \frac{j+3}{2} \binom{z+1}{j+2} (V_M^{(1)})^j},
\end{equation*}
which we can reduce with application of the binomial theorem and its derivative
to
\begin{equation}
V\mapsto g_{\gamma,z}(V) = \frac{\gamma}{1+\gamma}\frac{(Vz-1)(1+V)^z+(1+Vz)(1-V)^z}{V(z+1)\left[(1+V)^z - (1-V)^z\right]}.
\label{eq:Vrecur}
\end{equation}

\subsection{$x$-polarization in the isotropic system}
We note that, in the isotropic case, we should be able to derive the behavior
with $+x$ boundary conditions from the behavior with $+z$ boundary conditions.
In particular we can apply to every degree of freedom the $SU(2)$
Hadamard transformation
\[ H = \frac{1}{\sqrt{2}}\left(\begin{array}{c c}1&1\\1&-1\end{array}\right). \]
As an object that transforms under $SU(2)$, the factors that we have been
working with combine into a transfer matrix
\[ \rho_M = \left(\begin{array}{c c}Z_M&X^+_M\\ X^-_M&Y_M\end{array}\right). \]

Then we relate the $z$-polarized conditions, $X^-=X^+=0$, to the
$x$-polarized conditions, $X^-=X^+,Y=Z$, via
\[
H\left(\begin{array}{c c}Z&0\\0&Y\end{array}\right)H^\dagger
= \left(\begin{array}{c c}\frac{Y+Z}{2}&\frac{Z-Y}{2}\\\frac{Z-Y}{2}&\frac{Y+Z}{2}\end{array}\right),
\]
which allows us to make the substitution
\begin{equation}
V \leftarrow \frac{1-W}{1+W}.
\end{equation}

We can rewrite \eqref{eq:fcont} in more closed form as
\[ f_{z,\gamma}(W) = \frac{(1+\gamma z) - \gamma(z+1)W - (z+1)W^z + (z+\gamma)W^{z+1}}{(z+\gamma) - (z+1)W - \gamma(z+1)W^z + (1+\gamma z)W^{z+1}}. \]
Substituting $W\to\frac{1-V}{1+V}$,
\begin{align}
\tilde{g}_{z,\gamma}(V) \equiv \frac{1-f_{\gamma,z}(\sfrac{V+1}{V-1})}{1+f_{\gamma,z}(\sfrac{V+1}{V-1})}
= \frac{1-\gamma}{1+\gamma}\frac{1}{1+z}\frac{(1-zV)(1+V)^z-(1+zV)(1-V)^z}{V(1+V)^z-V(1-V)^z}
\label{eq:fhadamard}
\end{align}
When we compare \eqref{eq:Vrecur} with \eqref{eq:fhadamard}, we find that
the behavior of the deformed-decorated system with respect to the basal plane
matches the behavior of the same system with respect to the symmetry axis
for $\hat{\gamma}$ when
\begin{align}
\frac{\gamma}{1+\gamma} &= \frac{1-\hat\gamma}{1+\hat\gamma}\text{, or}\notag\\
\hat{\gamma} &= \frac{1}{1+2\gamma}.%\text{ and }\gamma=\frac{1}{2\hat\gamma}-\frac{1}{2}.
\label{eq:xz-duality}
\end{align}

Of course, in the undeformed case $\gamma=\sfrac{1}{2}$, the system is
isotropic, $\hat\gamma=\gamma$. In particular, for $z=10$ we see an
order/disorder transition in both the basal plane and the symmetry axis
at this value. More generally, however, we find that there are two transitions,
$\gamma_x$ and $\gamma_z$\footnote{Until now we have called $\gamma_z$
$\gamma_c$ instead.}, and
\begin{equation}
\gamma_x = \frac{1}{2\hat\gamma_z}-\frac{1}{2} = \frac{3}{z-4}.
\end{equation}
Note as well that this $xy$/$z$ ``duality'' \eqref{eq:xz-duality} maps the
entire accessible\footnote{Due to how $\gamma$ appears in the deformation, any
phase information it contains can be erased by a local unitary transformation;
alternatively, we may say that the formulas we have used $\gamma$ in should
more properly have had $|\gamma|$.} range $[0,\infty]$ $\gamma$ to
$\hat\gamma\in [0,1]$, meaning that these systems will not spontaneously order
antiferromagnetically along the basal plane.
\todo{Am I using these words right}

We therefore conclude that these systems have the following behaviors:
\todo[inline]{Maybe just replace all of this with a figure summarizing it}
\begin{itemize}
\item For $z=3$ there is no spontaneous symmetry breaking;
in all cases the system
should be disordered throughout both the ``ferromagnetic'' region
$\gamma \in [0,1)$ and the ``antiferromagnetic'' region $\gamma\in(1,\infty]$.
\item For $z=4$, when $\gamma_z=0$, we will observe critical behavior in the
basal plane only at the dual point $\gamma\to\infty$.
\item For $4<z<7$, we have an transition to order in the basal plane
in the $z$-antiferromagnetic region; in both such cases
$\gamma_x<\gamma_{Az}\equiv\gamma_z^{-1}$, dividing this region into
\begin{itemize}
  \item a fully disordered regime $1<\gamma<\gamma_x$,
  \item a $z$-disordered/$xy$-ordered regime $\gamma_x<\gamma<\gamma_{Az}$, and
  \item a fully ordered regime $\gamma>\gamma_{Az}$,
\end{itemize}
whereas the basal plane is disordered throughout the 
``ferromagnetic'' region $\gamma<1$.
\item For $z=7$, the transition into  will occur at the
``decoupling'' point $\gamma=1$; thus the basal plane will be
disordered throughout the ``ferromagnetic'' region and 
will obtain ferromagnetic order throughout the ``antiferromagnetic'' region.
\item For $7<z<10$, we will have $\gamma_z<\gamma_x<1$, such that, within
the antiferromagnetic region, there is
\begin{itemize}
\item an easy-axis regime (with $z$ order and
no $x$ order) $\gamma\in[0,\gamma_z)$,
\item a fully-disordered regime $\gamma\in(\gamma_z,\gamma_x)$,
\item and an easy-plane regime (with $x$ order and no $z$ order)
$\gamma\in(\gamma_x,1)$.
\end{itemize}
\item For $z=10$, when $\gamma_x=\gamma_z=\sfrac{1}{2}$, there is only symmetry-axis
order when $\gamma>\sfrac{1}{2}$ and only basal-plane order when
$\sfrac{1}{2}<\gamma<1$.
\item For $z>10$, $\gamma_x<\gamma_z$ and so there will be
\begin{itemize}
\item a phase where the symmetry axis spontaneously orders and the basal plane
does not, with $\gamma\in [0,\gamma_x)$,
\item a phase with spontaneous symmetry breaking in both the symmetry axis and
the basal plane for $\gamma \in (\gamma_x,\gamma_z)$ (it is easy to reason that we will have an easy axis for $\gamma<\sfrac{1}{2}$ and an easy plane for
$\gamma>\sfrac{1}{2}$; we will indeed see that this is the case), and
\item a phase where the basal plane spontaneously orders and the symmetry
axis does not, for $\gamma\in (\gamma_z,1)$.
\end{itemize}
\end{itemize}

We explore these phase diagrams for $z\in [8,12]$ in
Fig.~\ref{fig:phases-anisotropic}.

In the latter case, where $z>10$ and spontaneous symmetry breaking occurs in
both the symmetry axis and the basal plane for the region $(\gamma_x,\gamma_z)$,
it is worth exploring the behavior of the system some more\todo[disable]{rephrase}.
As noted in subsection~\ref{sec:loop}, we can define an infinite ground state
whenever a fixed point for our recursion relations exist; in particular, there
will be a family of $+z$-polarized ground states for $\gamma<\gamma_z$ and
a family of e.g. $+x$-polarized ground states for $\gamma>\gamma_x$. In
Fig.~\ref{fig:phases-anisotropic} we see evidence that in the ``hard-plane''
case $\gamma\in (\gamma_x,\sfrac{1}{2})$, $+x$ polarization is unstable, and
numerical error compounded by a recursive method creates 
polarization along the symmetry axis. Meanwhile, in the stable ``easy-axis'' and ``easy-plane''
cases ($+z$ polarization for $\gamma\in(0,\sfrac{1}{2})$ and $+x$ polarization for
$\gamma\in(\sfrac{1}{2},1)$), we have a family of ground states for a
Hamiltonian that is known to have a phase transition within the region in
question. However, we are unable to find any evidence that the one state
``sees'' the transition in the other direction, i.e. any nonanalyticities in the
$+z$-polarized state at $\gamma_x$ or in the $+x$-polarized state at $\gamma_z$.
\todo[inline]{Show loop simulations, or discuss
for hard-axis case.}

\section{Discussion}
We have analyzed AKLT systems on the Bethe lattice with an arbitrary number of
decorations on each edge. Our results -- for example, that
AKLT systems on singly-decorated Bethe lattice do not exhibit order for
$z\leq 10$ -- complement previous works which have shown that it is
easier to establish and lower-bound the gap of such systems
\cite{decorated1,decorated2,guo2020aklt,pomata2020degree3}  in that our
results imply that these systems are significantly further from
order or criticality than the corresponding systems on the ``bare'' lattice.

Based on our analysis, we have then developed AKLT-based
models with behavior exactly determined by recurrence relations.
In one of these systems, in which we used frustration to tune the strength of
interactions between sites without breaking $SU(2)$ symmetry, we found a
continuous phase transition for the $z>4$ cases in which the AKLT state on the
undecorated Bethe lattice is antiferromagnetically ordered, and exactly
determined critical exponents $\xi\sim \Delta\gamma^{-1}$,
$m \sim \Delta\gamma^{-\sfrac{1}{2}}$, and $S\sim r^{-\sfrac{1}{2}}$ for
$S$ a combination of observables designed to observe boundary effects at a
radius $r$. This suggests that similar decorations may be used
to perturb more physical ordered
frustration-free systems, such as the AKLT state on the cubic
lattice\cite{parameswaran2009order}, to phase transitions.
\todo[inline]{What about non-frustration-free systems?}

In another, in which we began with a singly-decorated Bethe lattice and then
deformed the decoration to obtain phase transitions, we saw competing
order in the symmetry axis and the basal plane, including coexisting ground
states with each order. It is notable, if not surprising given the definition
of frustration-freeness, that these states are all simultaneously true ground
states, with the same energy. We may ask if this is more generally the case for
frustration-free systems: if, when a (frustration-freeness-preserving)
deformation explicitly adds anisotropy in this way to a system that already
exhibits spontaneously-broken symmetry, it is generally possible to define
states in the thermodynamic limit that order along both ``easy'' and ``hard''
axes -- and, if so, whether there is a continuous phase transition that
is effectively undetectable by manipulations of a stable ground state.

\section*{Acknowledgements}
I would like to thank Tzu-Chieh Wei for suggesting the topic and for
fruitful discussions while the ideas in this work were being developed.
This work was supported by the National Science Foundation under
Grant No. PHY 1915165.

\bibliography{references.bib}

%apsrev4-2.bst 2019-01-14 (MD) hand-edited version of apsrev4-1.bst
%Control: key (0)
%Control: author (8) initials jnrlst
%Control: editor formatted (1) identically to author
%Control: production of article title (0) allowed
%Control: page (0) single
%Control: year (1) truncated
%Control: production of eprint (0) enabled
\begin{thebibliography}{14}%
\makeatletter
\providecommand \@ifxundefined [1]{%
 \@ifx{#1\undefined}
}%
\providecommand \@ifnum [1]{%
 \ifnum #1\expandafter \@firstoftwo
 \else \expandafter \@secondoftwo
 \fi
}%
\providecommand \@ifx [1]{%
 \ifx #1\expandafter \@firstoftwo
 \else \expandafter \@secondoftwo
 \fi
}%
\providecommand \natexlab [1]{#1}%
\providecommand \enquote  [1]{``#1''}%
\providecommand \bibnamefont  [1]{#1}%
\providecommand \bibfnamefont [1]{#1}%
\providecommand \citenamefont [1]{#1}%
\providecommand \href@noop [0]{\@secondoftwo}%
\providecommand \href [0]{\begingroup \@sanitize@url \@href}%
\providecommand \@href[1]{\@@startlink{#1}\@@href}%
\providecommand \@@href[1]{\endgroup#1\@@endlink}%
\providecommand \@sanitize@url [0]{\catcode `\\12\catcode `\$12\catcode
  `\&12\catcode `\#12\catcode `\^12\catcode `\_12\catcode `\%12\relax}%
\providecommand \@@startlink[1]{}%
\providecommand \@@endlink[0]{}%
\providecommand \url  [0]{\begingroup\@sanitize@url \@url }%
\providecommand \@url [1]{\endgroup\@href {#1}{\urlprefix }}%
\providecommand \urlprefix  [0]{URL }%
\providecommand \Eprint [0]{\href }%
\providecommand \doibase [0]{https://doi.org/}%
\providecommand \selectlanguage [0]{\@gobble}%
\providecommand \bibinfo  [0]{\@secondoftwo}%
\providecommand \bibfield  [0]{\@secondoftwo}%
\providecommand \translation [1]{[#1]}%
\providecommand \BibitemOpen [0]{}%
\providecommand \bibitemStop [0]{}%
\providecommand \bibitemNoStop [0]{.\EOS\space}%
\providecommand \EOS [0]{\spacefactor3000\relax}%
\providecommand \BibitemShut  [1]{\csname bibitem#1\endcsname}%
\let\auto@bib@innerbib\@empty
%</preamble>
\bibitem [{\citenamefont {Affleck}\ \emph {et~al.}(1987)\citenamefont
  {Affleck}, \citenamefont {Kennedy}, \citenamefont {Lieb},\ and\ \citenamefont
  {Tasaki}}]{AKLT_PRL}%
  \BibitemOpen
  \bibfield  {author} {\bibinfo {author} {\bibfnamefont {I.}~\bibnamefont
  {Affleck}}, \bibinfo {author} {\bibfnamefont {T.}~\bibnamefont {Kennedy}},
  \bibinfo {author} {\bibfnamefont {E.~H.}\ \bibnamefont {Lieb}},\ and\
  \bibinfo {author} {\bibfnamefont {H.}~\bibnamefont {Tasaki}},\ }\bibfield
  {title} {\bibinfo {title} {Rigorous results on valence-bond ground states in
  antiferromagnets},\ }\href {https://doi.org/10.1103/PhysRevLett.59.799}
  {\bibfield  {journal} {\bibinfo  {journal} {Phys. Rev. Lett.}\ }\textbf
  {\bibinfo {volume} {59}},\ \bibinfo {pages} {799} (\bibinfo {year}
  {1987})}\BibitemShut {NoStop}%
\bibitem [{\citenamefont {Affleck}\ \emph {et~al.}(1988)\citenamefont
  {Affleck}, \citenamefont {Kennedy}, \citenamefont {Lieb},\ and\ \citenamefont
  {Tasaki}}]{AKLT_1988}%
  \BibitemOpen
  \bibfield  {author} {\bibinfo {author} {\bibfnamefont {I.}~\bibnamefont
  {Affleck}}, \bibinfo {author} {\bibfnamefont {T.}~\bibnamefont {Kennedy}},
  \bibinfo {author} {\bibfnamefont {E.~H.}\ \bibnamefont {Lieb}},\ and\
  \bibinfo {author} {\bibfnamefont {H.}~\bibnamefont {Tasaki}},\ }\bibfield
  {title} {\bibinfo {title} {Valence bond ground states in isotropic quantum
  antiferromagnets},\ }\href {https://doi.org/10.1007/BF01218021} {\bibfield
  {journal} {\bibinfo  {journal} {Communications in Mathematical Physics}\
  }\textbf {\bibinfo {volume} {115}},\ \bibinfo {pages} {477} (\bibinfo {year}
  {1988})}\BibitemShut {NoStop}%
\bibitem [{\citenamefont {Fannes}\ \emph {et~al.}(1992)\citenamefont {Fannes},
  \citenamefont {Nachtergaele},\ and\ \citenamefont
  {Werner}}]{fannes1992cayley}%
  \BibitemOpen
  \bibfield  {author} {\bibinfo {author} {\bibfnamefont {M.}~\bibnamefont
  {Fannes}}, \bibinfo {author} {\bibfnamefont {B.}~\bibnamefont
  {Nachtergaele}},\ and\ \bibinfo {author} {\bibfnamefont {R.~F.}\ \bibnamefont
  {Werner}},\ }\bibfield  {title} {\bibinfo {title} {Ground states of vbs
  models on cayley trees},\ }\href@noop {} {\bibfield  {journal} {\bibinfo
  {journal} {Journal of statistical physics}\ }\textbf {\bibinfo {volume}
  {66}},\ \bibinfo {pages} {939} (\bibinfo {year} {1992})}\BibitemShut
  {NoStop}%
\bibitem [{\citenamefont {Werner}(1989)}]{werner1989construction}%
  \BibitemOpen
  \bibfield  {author} {\bibinfo {author} {\bibfnamefont {R.}~\bibnamefont
  {Werner}},\ }\bibfield  {title} {\bibinfo {title} {Construction and study of
  exact ground states for a class of quantum antiferromagnets},\ }\href@noop {}
  {\bibfield  {journal} {\bibinfo  {journal} {Revista Brasileira de
  F{\'\i}sica}\ }\textbf {\bibinfo {volume} {19}} (\bibinfo {year}
  {1989})}\BibitemShut {NoStop}%
\bibitem [{\citenamefont {Nachtergaele}(1990)}]{nachtergaele1990qmarkov}%
  \BibitemOpen
  \bibfield  {author} {\bibinfo {author} {\bibfnamefont {B.}~\bibnamefont
  {Nachtergaele}},\ }\bibfield  {title} {\bibinfo {title} {Working with quantum
  markov states and their classical analogues},\ }in\ \href@noop {} {\emph
  {\bibinfo {booktitle} {Quantum probability and applications V}}}\ (\bibinfo
  {publisher} {Springer},\ \bibinfo {year} {1990})\ pp.\ \bibinfo {pages}
  {267--285}\BibitemShut {NoStop}%
\bibitem [{\citenamefont {Fidaleo}\ and\ \citenamefont
  {Mukhamedov}(2005)}]{fidaleo-mukhamedov2005factors}%
  \BibitemOpen
  \bibfield  {author} {\bibinfo {author} {\bibfnamefont {F.}~\bibnamefont
  {Fidaleo}}\ and\ \bibinfo {author} {\bibfnamefont {F.}~\bibnamefont
  {Mukhamedov}},\ }\bibfield  {title} {\bibinfo {title} {On factors associated
  with quantum markov states corresponding to nearest neighbor models on a
  cayley tree},\ }in\ \href@noop {} {\emph {\bibinfo {booktitle} {Quantum
  Probability And Infinite Dimensional Analysis}}}\ (\bibinfo  {publisher}
  {World Scientific},\ \bibinfo {year} {2005})\ pp.\ \bibinfo {pages}
  {237--251}\BibitemShut {NoStop}%
\bibitem [{\citenamefont {Accardi}\ \emph {et~al.}(2016)\citenamefont
  {Accardi}, \citenamefont {Mukhamedov},\ and\ \citenamefont
  {Souissi}}]{accardi-mukhamedov2016construction}%
  \BibitemOpen
  \bibfield  {author} {\bibinfo {author} {\bibfnamefont {L.}~\bibnamefont
  {Accardi}}, \bibinfo {author} {\bibfnamefont {F.}~\bibnamefont
  {Mukhamedov}},\ and\ \bibinfo {author} {\bibfnamefont {A.}~\bibnamefont
  {Souissi}},\ }\bibfield  {title} {\bibinfo {title} {On construction of
  quantum markov chains on cayley trees},\ }in\ \href@noop {} {\emph {\bibinfo
  {booktitle} {Journal of Physics: Conference Series}}},\ Vol.\ \bibinfo
  {volume} {697}\ (\bibinfo {organization} {IOP Publishing},\ \bibinfo {year}
  {2016})\ p.\ \bibinfo {pages} {012018}\BibitemShut {NoStop}%
\bibitem [{\citenamefont {Friedman}(1997)}]{friedman1997dmrg}%
  \BibitemOpen
  \bibfield  {author} {\bibinfo {author} {\bibfnamefont {B.}~\bibnamefont
  {Friedman}},\ }\bibfield  {title} {\bibinfo {title} {A density matrix
  renormalization group approach to interacting quantum systems on cayley
  trees},\ }\href@noop {} {\bibfield  {journal} {\bibinfo  {journal} {Journal
  of Physics: Condensed Matter}\ }\textbf {\bibinfo {volume} {9}},\ \bibinfo
  {pages} {9021} (\bibinfo {year} {1997})}\BibitemShut {NoStop}%
\bibitem [{\citenamefont {{Nagaj}}\ \emph {et~al.}(2008)\citenamefont
  {{Nagaj}}, \citenamefont {{Farhi}}, \citenamefont {{Goldstone}},
  \citenamefont {{Shor}},\ and\ \citenamefont {{Sylvester}}}]{nagaj2008mps}%
  \BibitemOpen
  \bibfield  {author} {\bibinfo {author} {\bibfnamefont {D.}~\bibnamefont
  {{Nagaj}}}, \bibinfo {author} {\bibfnamefont {E.}~\bibnamefont {{Farhi}}},
  \bibinfo {author} {\bibfnamefont {J.}~\bibnamefont {{Goldstone}}}, \bibinfo
  {author} {\bibfnamefont {P.}~\bibnamefont {{Shor}}},\ and\ \bibinfo {author}
  {\bibfnamefont {I.}~\bibnamefont {{Sylvester}}},\ }\bibfield  {title}
  {\bibinfo {title} {{Quantum transverse-field Ising model on an infinite tree
  from matrix product states}},\ }\href
  {https://doi.org/10.1103/PhysRevB.77.214431} {\bibfield  {journal} {\bibinfo
  {journal} {\prb}\ }\textbf {\bibinfo {volume} {77}},\ \bibinfo {eid} {214431}
  (\bibinfo {year} {2008})},\ \Eprint {https://arxiv.org/abs/0712.1806}
  {arXiv:0712.1806 [cond-mat.stat-mech]} \BibitemShut {NoStop}%
\bibitem [{\citenamefont {{Abdul-Rahman}}\ \emph {et~al.}(2020)\citenamefont
  {{Abdul-Rahman}}, \citenamefont {{Lemm}}, \citenamefont {{Lucia}},
  \citenamefont {{Nachtergaele}},\ and\ \citenamefont {{Young}}}]{decorated1}%
  \BibitemOpen
  \bibfield  {author} {\bibinfo {author} {\bibfnamefont {H.}~\bibnamefont
  {{Abdul-Rahman}}}, \bibinfo {author} {\bibfnamefont {M.}~\bibnamefont
  {{Lemm}}}, \bibinfo {author} {\bibfnamefont {A.}~\bibnamefont {{Lucia}}},
  \bibinfo {author} {\bibfnamefont {B.}~\bibnamefont {{Nachtergaele}}},\ and\
  \bibinfo {author} {\bibfnamefont {A.}~\bibnamefont {{Young}}},\ }\bibfield
  {title} {\bibinfo {title} {{A class of two-dimensional AKLT models with a
  gap}},\ }in\ \href@noop {} {\emph {\bibinfo {booktitle} {Contemporary
  Mathematics}}},\ Vol.\ \bibinfo {volume} {741},\ \bibinfo {editor} {edited
  by\ \bibinfo {editor} {\bibfnamefont {H.}~\bibnamefont {Abdul-Rahman}},
  \bibinfo {editor} {\bibfnamefont {R.}~\bibnamefont {Sims}},\ and\ \bibinfo
  {editor} {\bibfnamefont {A.}~\bibnamefont {Young}}}\ (\bibinfo  {publisher}
  {American Mathematical Society},\ \bibinfo {address} {Providence},\ \bibinfo
  {year} {2020})\ pp.\ \bibinfo {pages} {1--21}\BibitemShut {NoStop}%
\bibitem [{\citenamefont {{Pomata}}\ and\ \citenamefont
  {{Wei}}(2019)}]{decorated2}%
  \BibitemOpen
  \bibfield  {author} {\bibinfo {author} {\bibfnamefont {N.}~\bibnamefont
  {{Pomata}}}\ and\ \bibinfo {author} {\bibfnamefont {T.-C.}\ \bibnamefont
  {{Wei}}},\ }\bibfield  {title} {\bibinfo {title} {{AKLT models on decorated
  square lattices are gapped}},\ }\href@noop {} {\bibfield  {journal} {\bibinfo
   {journal} {\prb}\ }\textbf {\bibinfo {volume} {100}},\ \bibinfo {eid}
  {094429} (\bibinfo {year} {2019})}\BibitemShut {NoStop}%
\bibitem [{\citenamefont {Guo}\ \emph {et~al.}(2020)\citenamefont {Guo},
  \citenamefont {Pomata},\ and\ \citenamefont {Wei}}]{guo2020aklt}%
  \BibitemOpen
  \bibfield  {author} {\bibinfo {author} {\bibfnamefont {W.}~\bibnamefont
  {Guo}}, \bibinfo {author} {\bibfnamefont {N.}~\bibnamefont {Pomata}},\ and\
  \bibinfo {author} {\bibfnamefont {T.-C.}\ \bibnamefont {Wei}},\ }\href@noop
  {} {\bibinfo {title} {The {AKLT} models on the singly decorated diamond
  lattice and two degree-4 planar lattices are gapped}} (\bibinfo {year}
  {2020}),\ \Eprint {https://arxiv.org/abs/2010.03137} {arXiv:2010.03137
  [cond-mat.str-el]} \BibitemShut {NoStop}%
\bibitem [{\citenamefont {Pomata}\ and\ \citenamefont
  {Wei}(2020)}]{pomata2020degree3}%
  \BibitemOpen
  \bibfield  {author} {\bibinfo {author} {\bibfnamefont {N.}~\bibnamefont
  {Pomata}}\ and\ \bibinfo {author} {\bibfnamefont {T.-C.}\ \bibnamefont
  {Wei}},\ }\bibfield  {title} {\bibinfo {title} {Demonstrating the
  {Affleck-Kennedy-Lieb-Tasaki} spectral gap on {2D} degree-3 lattices},\
  }\href {https://doi.org/10.1103/PhysRevLett.124.177203} {\bibfield  {journal}
  {\bibinfo  {journal} {Phys. Rev. Lett.}\ }\textbf {\bibinfo {volume} {124}},\
  \bibinfo {pages} {177203} (\bibinfo {year} {2020})}\BibitemShut {NoStop}%
\bibitem [{\citenamefont {Parameswaran}\ \emph {et~al.}(2009)\citenamefont
  {Parameswaran}, \citenamefont {Sondhi},\ and\ \citenamefont
  {Arovas}}]{parameswaran2009order}%
  \BibitemOpen
  \bibfield  {author} {\bibinfo {author} {\bibfnamefont {S.~A.}\ \bibnamefont
  {Parameswaran}}, \bibinfo {author} {\bibfnamefont {S.~L.}\ \bibnamefont
  {Sondhi}},\ and\ \bibinfo {author} {\bibfnamefont {D.~P.}\ \bibnamefont
  {Arovas}},\ }\bibfield  {title} {\bibinfo {title} {Order and disorder in
  {AKLT} antiferromagnets in three dimensions},\ }\href
  {https://doi.org/10.1103/PhysRevB.79.024408} {\bibfield  {journal} {\bibinfo
  {journal} {Phys. Rev. B}\ }\textbf {\bibinfo {volume} {79}},\ \bibinfo
  {pages} {024408} (\bibinfo {year} {2009})}\BibitemShut {NoStop}%
\end{thebibliography}%
\end{document}